\documentclass{AM}
\usepackage{float}
\usepackage{dcolumn,graphicx,color,booktabs,microtype,afterpage}
\usepackage{amssymb}
\usepackage{amsmath}
\usepackage[charter,greekuppercase=italicized]{mathdesign}
\usepackage[charter]{mathdesign}
\usepackage{sidecap}
\usepackage[mathlines]{lineno}
\usepackage[numbers,sort&compress]{natbib}
\usepackage{}
\graphicspath{{./}{figure/}}
\renewcommand{\tablename}{Table}
\makeatletter\renewcommand{\fnum@figure}[1]{\figurename~\thefigure.~}\makeatother
\makeatletter\renewcommand{\fnum@table}[1]{\tablename~\thetable.}\makeatother

\newcount\hh \newcount\mm
\hh=\time \divide\hh by 60
\mm=\hh \multiply\mm by 60 \mm=-\mm
\advance\mm by \time
\def\now{\number\hh:\ifnum\mm<10{}0\fi\number\mm}

\usepackage[colorlinks,plainpages=false,linkcolor=blue,urlcolor=blue,citecolor=blue,pdfpagemode=UseNone,pdfstartview=FitBH]{hyperref}
\UseRawInputEncoding

\begin{document}
	
\pagestyle{fancy}
\rhead{\includegraphics[width=2.5cm]{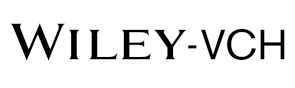}}	
%
%\title{Topological Hall effect far above room temperature\\
%in %\tcr{an} % If using singular (film)
%epitaxial Fe$_3$Ga$_4$ films}
%	
\title{Discovery of nodal-line superconductivity in chiral crystals}

\maketitle

\author{Tian Shang*}
\author{Jianzhou Zhao*}
\author{Lun-Hui Hu*}
\author{Weikang Wu}
\author{Keqi Xia}
\author{Mukkattu O. Ajeesh}
\author{Michael Nicklas}
\author{Yang Xu}
\author{Qingfeng Zhan}
\author{Dariusz J. Gawryluk}
\author{Ming Shi}
\author{Toni Shiroka}

% Dedication
%\dedication{Optional dedication here. If no dedication is required, please leave blank}
%\dedication{\dag These authors contributed equally to this work}

%

% Affiliations: Please provide adacemic titles (Prof. or Dr.) for all authors where applicable, and include an institutional email address for all corresponding authors
\begin{affiliations}
	%A. N. Author, A. N. O. Author\\
	%Address\\
	%Email Address:
	
	%A. N. O. Author\\
	%Address	

\vspace{1mm}
Prof. T. Shang\\
Key Laboratory of Polar Materials and Devices (MOE), School of Physics and Electronic Science, East China Normal University, Shanghai 200241, China\\
Chongqing Key Laboratory of Precision Optics, Chongqing Institute of East China Normal University, Chongqing 401120, China\\
Email Address: tshang@phy.ecnu.edu.cn\\

\vspace{1mm}
Prof. J. Zhao\\
Department of Physics, School of Science, Tianjin University, Tianjin 300354, China\\
Co-Innovation Center for New Energetic Materials, Southwest University of Science and Technology, Mianyang 621010, China\\
Email Address: jzzhao@swust.edu.cn\\

\vspace{1mm}
Prof. L. Hu, Prof. M. Shi\\
Center for Correlated Matter and School of Physics, Zhejiang University, Hangzhou 310058, China\\
Email Address: lunhui@zju.edu.cn\\

\vspace{1mm}
Prof. W. Wu\\
Key Laboratory for Liquid-Solid Structural Evolution and Processing of Materials (MOE), Shandong University, Jinan 250061, China\\

\vspace{1mm}
K. Xia,  Prof. Y. Xu,  Prof. Q. Zhan \\
Key Laboratory of Polar Materials and Devices (MOE), School of Physics and Electronic Science, East China Normal University, Shanghai 200241, China\\

\vspace{1mm}
Dr. M. O. Ajeesh, Dr. M. Nicklas\\
Max Planck Institute for Chemical Physics of Solids, N\"{o}thnitzer Str. 40, 01187 Dresden, Germany\\

\vspace{1mm}
Dr. D. J. Gawryluk\\
PSI Center for Neutron and Muon Sciences CNM, 5232 Villigen PSI,  Switzerland\\

%\vspace{1mm}
%Prof. M. Shi\\
%Center for Correlated Matter and School of Physics, Zhejiang University, Hangzhou 310058, China\\

\vspace{1mm}
Dr. T. Shiroka\\
PSI Center for Neutron and Muon Sciences CNM, 5232 Villigen PSI, Switzerland\\
Laboratorium f\"ur Festk\"orperphysik, ETH Z\"urich, CH-8093 Z\"urich, Switzerland\\

\end{affiliations}

% Dedication


% Abstract should be written in the present tense and impersonal style (i.e., avoid we), and be at most 200 words long
\begin{abstract}
\justifying

%-----below is AM version------------
Chiral crystals, whose key feature is the structural handedness, host exotic quantum phenomena driven by the interplay of band topology, spin-orbit coupling (SOC), and electronic correlations. Due to the limited availability of suitable chiral-crystal materials, their unconventional superconductivity (SC) remains largely unexplored.
%The unconventional superconductivity (SC) remains largely unexplored due to the limited availability of suitable chiral-crystal materials. 
Here, we report the discovery of unconventional SC in the La(Rh,Ir)Si family of materials by combining muon-spin spectroscopy, band-structure calculations, and perturbation theory. This family, characterized by a double-helix chiral structure, hosts exotic multifold fermions that are absent in other topological chiral crystals. While LaRhSi behaves as a fully-gapped superconductor, the substitution of 4$d$-Rh by 5$d$-Ir significantly enhances the SOC and leads to the emergence of topological nodal-line SC in LaIrSi. The developed model shows that the nodal-line SC arises from an isotropic SOC with a specific strength. Such an exotic mechanism expands our conventional understanding of material candidates for unconventional SC, which typically rely on a significantly anisotropic SOC to promote the triplet pairing. Our work establishes a new type of phase diagram, which provides a comprehensive roadmap for identifying and engineering unconventional SC in chiral crystals. Furthermore, it calls for renewed investigations of unconventional SC in other widely studied superconductors with a chiral structure.

\end{abstract}
\clearpage

\justifying
\section{Introduction}
Chirality, a quality of asymmetry resulting from the absence of inversion, mirror or roto-inversion symmetries, is one of the pivotal concepts in various fields, including physics, chemistry, and biology~\cite{Siegel1998,Bradlyn2016,Chang2018,Fecher2022,Naaman2019,Wang2023}. In solid-state physics, chirality usually refers to two main categories: static and dynamic.
%is typically divided into two main categories: static and dynamic. 
Static chirality pertains to the geometric arrangement of atoms in a crystal lattice, specifically structural chirality, while dynamic chirality is related to the spin or electronic properties~\cite{Dryzun2012,Hasan2021,Narang2021,Kumar2021,Yan2023,Yang2024}. Electronic chirality is usually associated with the spin-momentum locking of particles or quasiparticles near degenerate points in momentum space (e.g., Weyl fermions)~\cite{Yan2017,Armitage2018,Nagaosa2020,Lv2021,Amato1997}. In contrast, spin (or magnetic) chirality characterizes the topological spin textures or domain walls in real space, such as magnetic skyrmions~\cite{Nagaosa2013,Fert2017,Yang2021,Tokura2021,Cheong2022,Zhang2025,An2023,Kim2024}. The combination
of structural and electronic chirality triggers the emergence of topological chiral crystals, which exhibit a plethora of exotic quantum phenomena. These include fermionic excitations characterized by large Chern numbers ($\lvert C \rvert >$ 1, with no analogues in high-energy physics), unique spin textures (e.g., parallel spin-momentum locking), giant helicoid Fermi arcs, a large topologically nontrivial energy window, chiral magnetic effects, and quantized circular photogalvanic effect~\cite{Yan2023,Hasan2021,Narang2021,Bradlyn2016,Chang2018,Tang2017,Lin2022,Fernando2017,Tan2022,Chang2020}. These chirality-induced effects %could offer
promise new directions in fundamental science and new technological applications in spintronics, optoelectronics, and biochemistry.

Typical examples of topological chiral crystals are the
cubic B20 materials with a noncentrosymmetric space group $P2_13$ (No.~198). Among these materials, the most prominent ones
are the transition-metal monosilicides or monoaluminides $T$$X$,
where $T$ represents a transition metal and $X$ denotes Si, Ge, Sn, Al, and Ga.
In particular, both CoSi and RhSi host a threefold spin-1 Weyl fermion with a topological charge $C$ = 2 at the center of the Brillouin zone (BZ) ($\Gamma$ point) and a fourfold charge-2 Weyl fermion with an opposite topological charge at the zone boundary ($R$ point)~\cite{Rao2019,Sanchez2019,Takane2019,Sanchez2023}. Recent photoemission spectroscopy studies have confirmed the presence of unconventional multifold chiral fermions close to the Fermi energy, along with their associated extremely long helicoid Fermi arcs traversing the entire surface BZ~\cite{Rao2019,Sanchez2019,Takane2019,Schroter2019,Schroter2020,Yao2020,Sessi2020,Sanchez2023}. Moreover, spin-orbit coupling (SOC) has been shown to induce a spin splitting of the topological band degeneracies, effectively doubling the topological charge from $\lvert C \rvert = 2$ to 4, as experimentally verified in PtAl, PtGa and PdGa~\cite{Schroter2019,Schroter2020,Yao2020,Sessi2020,Krieger2024}. Additionally, 
%In addition, 
these chiral crystals serve as %are also 
model systems for exploring topological bosonic excitations~\cite{Zhang2018,Miao2018}. 

Despite significant advancements in understanding their topological
aspects, the role of many-body interactions in topological chiral crystals, and how their band topology affects the electronic correlations, %effect 
remains largely unexplored and attracts increased attention. Notably, phenomena such as charge order and superconductivity (SC) have been reported in a few chiral crystals. The incommensurate charge-density-wave (CDW) order formed in the topological boundary states of the CoSi (001) surface is a striking example of chirality's impact
on the electronic states~\cite{Li2022,Rao2023}, which, however, is missing on the other surfaces or in the bulk. The orientation of the CDW phase can be tuned by the chirality of the Fermi arcs, i.e., by the structural chirality~\cite{Li2022,Rao2023}, and is closely related to the van Hove singularities~\cite{Sanchez2023}, the latter possibly inducing SC as well~\cite{Chang2018,Hasan2021,Mardanya2024}. So far, a limited number of chiral crystals has
been found to exhibit SC~\cite{Carnicom2018,Sun2015,yuan2006,nishiyama2007,Karki2010}, including some B20 chiral crystals,  such as AuBe~\cite{Amon2018,Khasanov2020b}, RhGe~\cite{Tsvyashchenko2016}, and BiPdSe~\cite{Joshi2015}. 
%~\cite{Amon2018,Rebar2019,Singh2019b,Beuer2019,Khasanov2020b,Khasanov2020,Datta2022,Mardanya2024,Tsvyashchenko2016,Joshi2015}. 
While most of them behave as conventional superconductors~\cite{Smidman2017},
%~\cite{Smidman2017,Amon2018,Rebar2019,Singh2019b,Beuer2019,Khasanov2020b,Khasanov2020,Datta2022},
only Li$_2$Pt$_3$B is known to exhibit spin-triplet pairing~\cite{yuan2006,nishiyama2007}, which also shows
a gapless topological SC with Majorana surface modes~\cite{Gao2022}.
%Similarly, while Li$_2$Pd$_3$B and Li$_2$Pt$_3$B show analogous topological properties to B20 chiral crystals~\cite{Bradlyn2016,Gao2022},
%\tcr{only the latter exhibits} %the latter shows
% This sentence is not very clear. Rerormulate if possible. TS
%a gapless topological SC with Majorana surface modes~\cite{Gao2022}.
It has recently been predicted that RhGe and PdBiSe host mixed-parity pairing and topological superconductivity, involving chiral fermions at time-reversal invariant momenta (TRIM)~\cite{Mardanya2024,Lv2019}. In addition, topological SC arising from a $s_+ \oplus s_-$ pairing has been theoretically investigated in RhSi~\cite{Lee2021}. Unfortunately,
the %microscopic 
superconducting nature of these chiral crystals has not yet been explored 
%remain largely unexplored 
due to the absence of SC, the difficulty in synthesizing a material with bulk SC~\cite{Tsvyashchenko2016,Kamaeva2022,Joshi2015}, and their extremely low $T_c$s ($<$ 0.2\,K)~\cite{Okamoto2020,Mizutani2019}.

Despite previous studies of SC in chiral crystals, the chirality-driven mechanisms
behind these novel superconducting states and the effect of SOC on
the superconducting pairing have not been explored. %understood. 
Here, we report
the discovery of unconventional SC in cubic La(Rh,Ir)Si chiral crystals and reveal the pivotal role of SOC in %determining their
the superconducting pairing. Similar to the CoSi family of chiral crystals, there is also a large number of unconventional multifold fermions in La(Rh,Ir)Si, which significantly affect the band topology and superconducting states. 
%have crucial impacts on the band topologies and superconducting states. 
Remarkably, while LaRhSi behaves as a fully-gapped superconductor, LaIrSi shows nodal-line SC. Contrary to the B20 transition-metal monosilicides or monoaluminides, the incorporation of La atoms introduces an additional chiral chain in La(Rh,Ir)Si, featuring a unique double-helix chiral structure. Density-functional-theory (DFT) calculations reveal that La(Rh,Ir)Si harbor more electronic bands near the Fermi energy than the CoSi family, pointing to a chirality-driven mechanism in the emergence of SC. Moreover, the substitution of 4$d$-Rh with 5$d$-Ir atoms enhances significantly the strength of SOC, culminating in a topological nodal-line superconducting state in LaIrSi. To understand how SC evolves %the evolution of SC
from nodal to nodeless as the SOC is reduced, we develop a phenomenological theory which reveals that nodal-line SC is stabilized by %both
an isotropic SOC with a well-defined strength, as well as a spin-independent $C_3$-wrapping term. Such an exotic mechanism
for establishing triplet pairing goes beyond the conventional ones. 
The latter typically rely on either a significantly anisotropic SOC (in superconductors without an inversion center), or strong spin fluctuations (in magnetic superconductors).  
The nodal-line SC in LaIrSi is also different from that in cuprates, the latter hosting $d$-wave spin-singlet pairs. 
Our theoretical model suggests that a significant SOC is not required %to be significant
in order to induce a nodal SC. Consequently, the new mechanism can bypass 
the restrictions normally imposed on the materials expected to exhibit unconventional SC.
The nodal-line phase diagram we propose here can serve as a preliminary guide
in the search for unconventional SC in other chiral crystals.

\section{Results and Discussion}
\begin{figure}
	\centering 
	\includegraphics[width=0.95\linewidth]{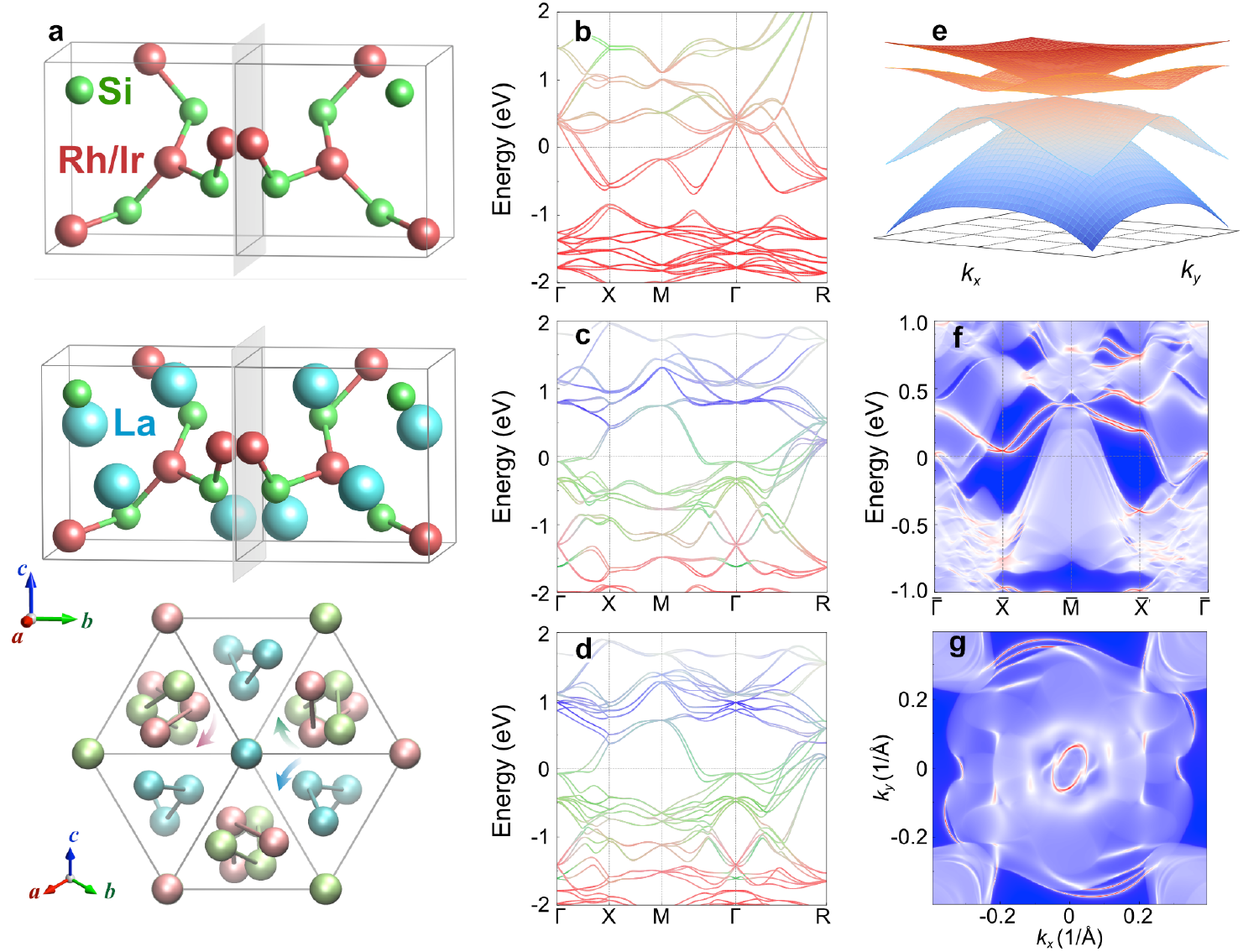}
	\caption{\label{fig:bands} Chiral crystal structure and electronic-band calculations.
	    a) Illustration of the chiral crystal structure of the (Rh,Ir)Si sublattice and La(Rh,Ir)Si. Both the right-handed and left-handed counterparts are presented. The double-helix arrangements are clearly visible in the top view
		along the [111]-direction (bottom panel). b-d) The calculated electronic band structure by considering the SOC for RhSi sublattice (b), LaRhSi (c), and LaIrSi (d). Note that the lattice constants and the atomic coordinates of the RhSi sublattice are different from those of the RhSi chiral crystal (see details in Table S1, Supporting Information).
		The Rh-$4d$/Ir-$5d$, La-$5d$, and Si-$3p$ orbitals are shown in red, blue, and green, respectively. The electronic bands of other isostructural compounds are shown in Figures~S10-S18 in the Supporting Information. e) Energy dispersion of the LaIrSi bands below the Fermi level in the $k_x$--$k_y$ plane around the high-symmetry $\Gamma$ point. f,g) The (001) surface states (f) and the surface Fermi arcs (g) of LaIrSi at the fourfold degenerate $\Gamma$ point with a topological charge $C = +2$.}
\end{figure}

\textbf{Double-helix chiral structure}.
A detailed knowledge of the crystal %lattice 
structure is crucial because the symmetry principles inherent in it
dictate the related phases within the framework of Landau theory~\cite{chaikin1995principles,dresselhaus2007group}.
As depicted in Figure~\ref{fig:bands}a, a chiral crystal surface
is characterized by an arrangement of atoms that does not 
coincide with its mirror image when reflected perpendicular to the surface plane. The binary AuBe alloy is a noncentrosymmetric superconductor
with a B20 chiral structure~\cite{Amon2018}, sharing the same space
group $P2_13$ (No.~198) with the CoSi family and La(Rh,Ir)Si,
the latter being the focus of the current work. 
%It is noted that AuBe is a noncentrosymmetric superconductor with a B20 chiral structure~\cite{Amon2018}, sharing the same space group $P2_13$ (No.~198) as LaRhSi and LaIrSi, the latter two being the focus of the current work. 
While AuBe is widely regarded as a conventional superconductor,
with a dominant %an isotropic gap indicative of predominantly 
$s$-wave singlet pairing in the weak-coupling limit~\cite{Khasanov2020b,Singh2019b}, we uncover unconventional SC in La(Rh,Ir)Si.
%However, in this work, we uncover unconventional superconductivity in LaRhSi and LaIrSi. 
Such a striking difference between AuBe and La(Rh,Ir)Si hints at an intimate interplay between the crystal structure and the electronic
correlations in materials with a chiral structure.   
%This striking inconsistency between AuBe and La(Rh,Ir)Si indicates a complex interplay between lattice structure and electronic correlations. 
Here, their cubic crystal structure contains %the symmetries of
a twofold screw axis and a threefold rotation axis. The CoSi family and AuBe exhibit a pair of sublattice distortions along the [111]-direction (i.e., the chiral axis). In the case of La(Rh,Ir)Si,
the structure consists of two sublattices. The (Rh,Ir)Si sublattice resembles that of AuBe and CoSi family (see details in Table~S1 in the Supporting Information), while the La sublattice forms a  separate chiral chain. To elucidate this, both structures of the
(Rh,Ir)Si sublattice and La(Rh,Ir)Si are depicted in Figure~\ref{fig:bands}a. 
%the crystal structure of La(Rh,Ir)Si is separated into (Rh,Ir)Si and La sublattices, the former is quite similar to the structure of CoSi family of materials (see details in Supplementary Table 1).     
%commence with an analysis of their crystal structures before proceeding to theoretical calculations. 
The Rh/Ir and the Si atoms exhibit a %crystallize
clockwise arrangement when viewed along the [111]-direction,
which presents a distinct structural chirality. Conversely, the La sublattice exhibits a chiral chain with an anticlockwise arrangement. 
%analogous to Rh or Ir atoms.
%As depicted in the middle panel of Fig.~\ref{fig:bands}a, the La atoms exhibit a chiral structure analogous to that of the Rh/Ir transition metals. 
Generally, the chirality can be distinguished by the handedness of the helix, 
%, such as that formed by the La atoms, % this is very confusing! TS
either clockwise or anticlockwise, depending on the enantiomer. %(Fig.~\ref{fig:bands}a). %\tcr{[LH-comment]: can we cite SM to add one sentence from experimental data?}
Considering that the (Rh,Ir)Si and La sublattices form chiral chains,
but with opposite chirality, the La(Rh,Ir)Si family of materials can be
classified as double-helix chiral crystals. 
Note that, to date, no SC has been found in the isostructural
RhSi chiral crystal. However, due to the presence of two chiral
sublattices, a double-helix chiral crystal may host a greater abundance
of electronic states. This can account for the appearance of unconventional superconducting phenomena in La(Rh,Ir)Si.   

%This leads to multiband superconductivity, holding promise for the realization of novel superconducting phenomena. Further investigation based on Density Functional Theory (DFT) calculations below will elucidate this.
%Both AuBe and La(Rh,Ir)Si possess cubic structures, featuring the symmetry of a \tcr{twofold} screw axis and a \tcr{threefold} rotation axis. AuBe exhibits a pair sublattice distortion along the [111] direction (chiral axis). 
%In the case of La(Rh,Ir)Si, the structure consists of two components: the (Rh,Ir)Si sublattice resembles that of AuBe, while additional La atoms form a separate chiral chain, as depicted in Fig.~\ref{fig:bands} (a). Therefore, we refer to this arrangement in La(Rh,Ir)Si as a double-helix chiral structure for comparative analysis. Additionally, it is noteworthy that the chiral material RuSi is not a superconductor itself. From this perspective, the double-helix materials may host a greater abundance of electronic states within their chiral sublattices. This leads to multiband superconductivity, holding promise for the realization of novel superconducting phenomena. Further investigation based on Density Functional Theory (DFT) calculations below will elucidate this. 

\textbf{Topological electronic band structure.} We first performed band-structure calculations via density functional theory to investigate the impact of the aforementioned double-helix chiral structure on the electronic properties of La(Rh,Ir)Si. 
%formation of superconducting states in La(Rh,Ir)Si, we performed the detailed electronic band structure calculations via density functional theory(DFT). 
The density of states (DOS) close to the Fermi level ($E_f$) is dominated by the La-$5d$, Rh-$4d$ (or Ir-$5d$), and Si-$3p$ orbitals (see Figure~S1, Supporting Information). 
In both LaRhSi and LaIrSi, the $d$-orbital of Rh/Ir and La atoms contributes over 22\% and 15\% of the total DOS, respectively, highlighting the important role of the La chiral chain. 
The electronic band structures of the RhSi sublattice, LaRhSi, and LaIrSi are shown in Figure~\ref{fig:bands}b-d. 
Similar to the CoSi family of chiral crystals, the electronic bands near the Fermi energy are dominated by the $d$ orbitals in the 
RhSi or IrSi sublattice (see band structures of B20 RhSi and IrSi chiral crystals in Figures~S2 and S3 in the Supporting Information).
%Interestingly, the band structures of the RhSi/IrSi sublattices and LaRhSi/LaIrSi are clearly distinct from those of the CoSi family of chiral crystals near the Fermi energy (see band structures of RhSi and IrSi chiral crystals in Supplementary Fig.~2 and 3).
%The electronic band structures of LaIrSi and its IrSi and La sublattices are shown in Fig.~\ref{fig:bands} b,c,d. The IrSi sublattice exhibits a band structure similar to that of the CoSi family of chiral crystals. 
Interestingly, despite sharing the same crystal structure,
near the Fermi energy the band structures of LaRhSi and LaIrSi are
distinct from that of the CoSi chiral family. %of chiral crystals
This confirms the key role of the La chiral chain in the
appearance of SC in these double-helix chiral materials. For example,
without considering SOC, RhSi or IrSi host a threefold band
crossing at the $\Gamma$ point, while a twofold and a threefold
band crossing can be identified in LaRhSi or LaIrSi near the Fermi energy. 
By comparing the band structures of LaRhSi and LaIrSi --- in
the presence and absence of SOC --- it turns out that SOC lifts
the band degeneracies, except along the $X$--$M$ direction (Figure~\ref{fig:bands}c,d and Figures.~S2 and S3, Supporting Information).
The band splitting $E_\mathrm{SOC}$ caused by the SOC at the high-symmetry $R$ point is about 207 and 37\,meV for LaIrSi and LaRhSi, respectively. %corresponding to $E_\mathrm{SOC}$/k$_\mathrm{B}T_c$ $\approx$ \tcr{1144}  and \tcr{98} \tcr{[Comment]: However, the value of $T_c$ is introduced below}, 
Such a significant SOC effect in LaIrSi is mainly attributed to the
5$d$ orbitals of the La and Ir atoms near $E_f$.
Compared to most other noncentrosymmetric superconductors, the
$E_\mathrm{SOC}$ of LaIrSi is much higher~\cite{Smidman2017}.
Up to 10 bands are identified to cross the Fermi level of both materials, confirming their multiband character. 

Due to their chiral structure, La(Rh,Ir)Si, similar to the CoSi family of materials~\cite{Rao2019,Sanchez2019,Takane2019,Schroter2019,Schroter2020,Yao2020,Sessi2020,Sanchez2023,Krieger2024}, exhibit a rich set of topological features. These give rise to multifold Weyl fermions at high-symmetry momenta and spin-polarized Fermi arc states spanning the entire surface Brillouin zone (Figure~\ref{fig:bands}e-g).
For instance, at the zone boundary ($R$ point), a double Weyl fermion arises in the absence of SOC. When SOC is included, this excitation splits into a spin-1 state and a single Weyl point.
Despite sharing similar symmetry constraints with the CoSi family, La(Rh,Ir)Si display distinct topological properties. 
In the absence of SOC, the band structure at the $\Gamma$ point hosts two kinds of topological excitations near the Fermi level, a twofold degeneracy with a topological charge of $C = +4$ and a threefold degeneracy with a topological charge of $C = -2$ (see Figures.~S2 and S3, Supporting Information).
%realizes a two fold degeneracy at $\Gamma$ point with topological charge of +4. 
%The band splitting induced by the SOC in CoSi family is relative small, which is less than 2~meV\cite{Wang2020,Xu2019}.
Introducing SOC lifts the twofold degeneracy at the $\Gamma$ point
and transforms it into a fourfold degeneracy with a topological
charge $C = +2$, a hallmark topological feature absent in the CoSi family.
Such distinct topological features arise from the additional
electronic bands peculiar to the double-helix chiral structure
of La(Rh,Ir)Si. 
%\tcr{[Comment]: I remenber there are also other ternary compounds exist such characters. Should we mention here?} 
%\tcb{LH: maybe cite some papers?}
The comparison of the band structures of La(Rh,Ir)Si
with those of B20 (Rh,Ir)Si chiral crystals is presented in
Figures.~S2 and S3 in the Supporting Information.

%indicates that the differences is arising from the double-helix chiral structure.} 
% \tcr{[Comment]: do we have a figure to support this conclusion? e.g., compare Bands of RhSi and those of LaRhSi? Jianzhou, maybe add such a figure in the SM}
Moreover, since the sum of topological charges in the Brillouin zone must equal zero, we identify a compensating topological charge around the $R$ point. This leads to a unique bulk-boundary correspondence, where the topological surface states form a continuous connection between the $\Gamma$ and $R$ points (or surface-projected $\bar{M}$ point). DFT calculations in La(Rh,Ir)Si
%As shown in Fig.~\ref{fig:bands}f,g, our calculations 
reveal spin-polarized helicoid Fermi arcs spanning their entire
surface Brillouin zone (Figure.~\ref{fig:bands}f,g).
Since in this case the multifold fermions at the high-symmetry
momentum points are very close to the Fermi level (i.e., $\sim 29$\,meV
below $E_f$ at the $\Gamma$ point), La(Rh,Ir)Si represent an ideal
system for investigating the relationship among the chirality-induced
band topology, SOC, and unconventional SC, as we discuss in detail below.

\begin{figure}
  \centering 
  \includegraphics[width=0.7\linewidth]{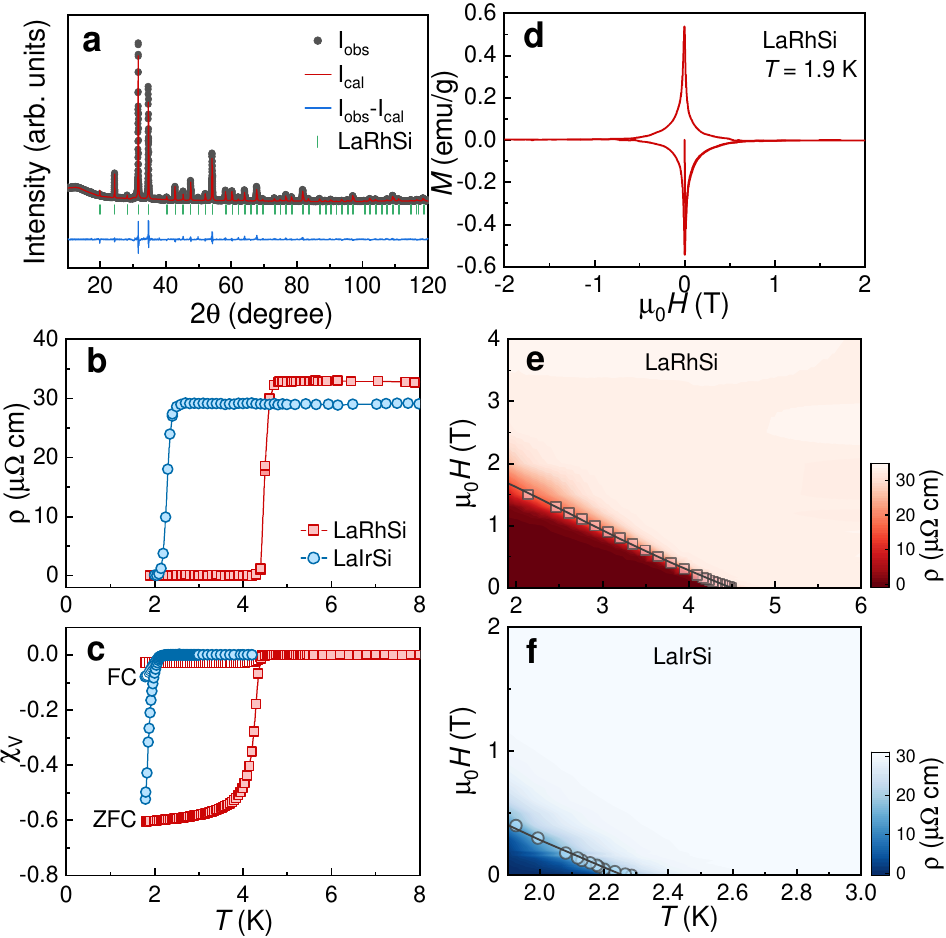}
  \caption{\label{fig:SC} Characterization of superconductivity. 
  	a) Room-temperature XRD pattern and Rietveld refinements for LaRhSi. The black circles and the solid red line represent the experimental pattern and the Rietveld refinement profile, respectively. The blue line at the bottom shows the residuals, i.e., the difference between the calculated and experimental data. The vertical bars mark the calculated Bragg-peak positions corresponding to the $P2_13$ space group. LaIrSi shows a similar XRD pattern. 
  	b,c) Temperature dependence of the zero-field electrical resistivity (b) and magnetic susceptibility (c) for LaRhSi and LaIrSi. The magnetic susceptibility data were collected using both field-cooled (FC) and zero-field-cooled (ZFC) protocols. 
  	d) Field-dependent magnetization $M(H)$ collected at 1.9\,K for LaRhSi. The $M(H)$ at other temperatures are presented in Figure S5 in the Supporting Information.        
  	e,f) Upper critical fields of LaRhSi (e) and LaIrSi (f). The symbols represent the middle of the superconducting transition (i.e., 50\%\,$\rho_\mathrm{n}$, where $\rho_\mathrm{n}$ is the normal-state resistivity at temperature just above $T_c$). Solid lines are fits to the Ginzburg-Landau model~\cite{Zhu2008}. The background color represents the magnitude of electrical resistivity $\rho(T,H)$. }
\end{figure}

\textbf{Characterization of superconductivity.} Although both LaRhSi and LaIrSi compounds were known to exhibit SC a few decades ago~\cite{Chevalier1982,Braun1984}, and the superconducting transition temperature $T_c$ was reported, their superconducting properties remain largely unexplored. We synthesized LaRhSi and LaIrSi using the arc-melting method~\cite{Chevalier1982} and characterized their SC via magnetic susceptibility and electrical resistivity measurements. Both compounds crystallize in a cubic chiral structure with double helices (Figure.~\ref{fig:bands}a), confirmed by % the
powder x-ray diffraction (XRD) refinements (Figure~\ref{fig:SC}a).
%The refined atomic coordinates are summarized in Supplementary Table~1.
Both LaRhSi and LaIrSi show metallic behavior in the studied temperature
range (Figure~S4, Supporting Information). The electrical resistivity in
Figure~\ref{fig:SC}b drops to zero at $T_c = 4.4$\,K and 2.1\,K for LaRhSi
and LaIrSi, respectively,
%For LaRhSi and LaIrSi, electrical resistivity in Fig.~\ref{fig:SC}b drops to zero at $T_c$ = 4.4\,K and 2.1\,K, respectively,
where the magnetic susceptibilities in Figure~\ref{fig:SC}c also show a clear diamagnetic response. % too. %Both $T_c$s are consistent with previous studies~\cite{Chevalier1982,Braun1984}. 
The well-separated ZFC- and FC-curves indicate a type-II SC, consistent with
the field-dependent magnetization shown in Figure~\ref{fig:SC}d.
The upper critical fields determined from the electrical resistivity measured under various magnetic fields (Figure~\ref{fig:SC}e,f) are $\mu_0$$H_\mathrm{c2}$(0) =  2.42(1) and 2.35(6)\,T for LaRhSi and LaIrSi, respectively (see details in Figure~S6 and Note~S1, Supporting Information). In addition, the positive curvature in the $H_\mathrm{c2}(T)$ data of both materials is most likely attributed to the multiband nature of their SC (Figure~S6, Supporting Information). While this is consistent with the electronic band structures
shown in Figure~\ref{fig:bands}c,d, it requires further investigation.

\textbf{{\textmu}SR and superconducting pairing.} To investigate the superconducting nature of LaRhSi and LaIrSi, 
%Though the transition temperature $T_c$ of LaRhSi and LaIrSi was reported~\cite{Chevalier1982,Braun1984}, their superconducting pairing have never been explored. 
we first performed zero-field (ZF)-{\textmu}SR to
establish if time-reversal symmetry (TRS) is spontaneously
broken in the superconducting state. 
ZF-{\textmu}SR represents one of the most highly sensitive techniques for detecting weak magnetic fields (down to $\sim$ 0.01 mT) owing to the large muon gyromagnetic ratio (851.615 MHz/T) and the availability of nearly 100\% spin-polarized muon beams~\cite{Amato1997,Yaouanc2011,Amato2024,Blundell2021}.
An increase in the zero-field muon-spin relaxation rate below the onset of SC provides direct evidence for a superconducting state with broken TRS~\cite{Shang2021b,Ghosh2020b}. In both LaRhSi and LaIrSi, the %practically
nearly overlapping ZF-{\textmu}SR spectra indicate the absence of a spontaneous field below $T_c$, 
confirming that TRS is preserved
in the superconducting state of both compounds (Figure~\ref{fig:muSR}a,b). Indeed, the derived muon-spin relaxation rates are almost identical in the superconducting and normal states, with their differences %lying 
falling within the standard deviations (see Note~S2, Supporting Information).

Next, we examine the SC pairing symmetries of LaRhSi and LaIrSi
by detecting the temperature-dependent magnetic penetration depth via
the transverse-field (TF) {\textmu}SR technique. Since the superfluid density $\rho_\mathrm{sc}$ is proportional to the inverse square of the magnetic penetration depth (i.e., $\rho_\mathrm{sc}$  $\propto  \lambda_\mathrm{eff}^{-2}$), the measured temperature-dependent $\lambda_\mathrm{eff}(T)$ reveals the nature of the superconducting pairing (see details in Notes~S2 and S3, Supporting Information)~\cite{Amato1997,Yaouanc2011,Amato2024,Blundell2021}.
Due to the formation of the flux-line lattice (FLL), the TF-{\textmu}SR
spectra collected in the superconducting state of both LaRhSi and
LaIrSi (Figure~\ref{fig:muSR}c,d) show enhanced muon-spin relaxation
rates, %that are
determined by the magnetic penetration depth. 
The in\-verse\--square of the effective magnetic penetration depth
$\lambda_\mathrm{eff}^{-2}(T)$ vs.\ the reduced temperature $T/T_c$ for LaRhSi and LaIrSi is shown in Figure~\ref{fig:muSR}e,f.
The two $\lambda_\mathrm{eff}^{-2}(T)$ evolutions are clearly distinct.  
In LaRhSi, the temperature-invariant superfluid density at
low temperatures suggests a fully-gapped superconducting state. By contrast, the strongly temperature-dependent superfluid density hints at the presence of low-energy excitations and thus, of gap nodes in LaIrSi. In this case,
$\lambda_\mathrm{eff}(T)$ shows a subquadratic $T^n$ ($n \sim 1.0$--1.5)
rather than a quadratic temperature dependence (see Figure~S7 and Note~S3, Supporting Information),
which is consistent with the presence of line nodes. %in LaIrSi. 
As shown by solid lines in Figure~\ref{fig:muSR}e,f, 
the $\rho_\mathrm{sc}(T)$ is well described by the proposed
theoretical model involving mixed singlet-triplet pairing (see details below). 
%Figures~\ref{fig:muSR}c and d show the representative TF-$\mu$SR spectra collected in the normal and superconducting states of LaRhSi and LaIrSi, respectively. The development of a flux-line lattice in the mixed state leads to an enhanced muon-spin relaxation rate that is determined by the magnetic penetration depth. 
%The small muon-spin relation rate in the superconducting state reflects the large magnetic penetration depth in LaRhSi and LaIrSi superconductors. 
%reflected by a large SOC in both LaRhSi and LaIrSi NCSCs (see details in the Discussion section). 
While the singlet channel is dominant in LaRhSi, in LaIrSi, due
to the significantly enhanced SOC (see Figure~\ref{fig:bands}
and Figures.~S2 and S3, Supporting Information), the triplet component is larger than
the singlet one, leading to the superconducting gap nodes.

\textbf{Theory of nodal superconductivity.} 
%\tcb{Shang: Shall we add a few sentences to mention the 198 chiral space group and the related physics?}
%Electrons exhibit both spin and orbital polarization as they pass through chiral crystals, the spin polarization is proportional to intrinsic SOC while the orbital polarization is insensitive to SOC~\cite{Yang2024}. Therefore, the SOC can tune the Cooper parings in different manner in the spin and orbit channel in the superconductors with a chiral structure.
%$\tcr{Morerover, the superconducting pairing involving multifold fermions in the chiral crystals may suggest the presence of nontrivial superconducting pairing symmetries. 
	%We proposed an underlying mechanism to explain the observed nodal SC in LaIrSi, which strongly relies on its significant SOC effect (see band-structure calculations in Fig.~\ref{fig:bands} and Supplementary Fig.3 and 4). 
%
In chiral crystals, the coexistence of SC with
chirality-induced multifold fermions %in chiral crystals
can stabilize unconventional pairing symmetries, as exemplified by the nodal
SC observed in LaIrSi. Our findings reveal that LaIrSi exhibits nodal SC with significantly stronger SOC than LaRhSi (see band-structure calculations in Figure~\ref{fig:bands} and Figures.~S2 and S3, Supporting Information), which instead remains fully gapped. This comparison highlights the crucial role of SOC in driving nodal SC in chiral crystals (see details also in Note~S4, Supporting Information).
In general, an enhanced SOC increases the degree of mixing between spin-singlet ($\Delta_s$) and spin-triplet ($\Delta_t$) components, a hallmark of noncentrosymmetric superconductors where broken inversion symmetry permits such a mixing~\cite{bauer2012,Smidman2017}. This mechanism has been %applied
invoked to explain nodal SC
	%naturally accounts for nodal gap structures 
in systems like Li$_2$Pt$_3$B~\cite{yuan2006,nishiyama2007}, CePt$_3$Si~\cite{Bauer2004}, and CaPtAs~\cite{Shang2020a}. It is important to note that previous approaches are based on a topologically trivial Hamiltonian with isotropic Fermi surfaces, while the spin-triplet component exhibits significant anisotropy~\cite{schnyder2015topological}. However, in a cubic lattice (e.g., LaIrSi), the leading-order spin-triplet pairing is constrained by the cubic symmetry to adopt an isotropic form, with a $d$-vector given by $d_{\boldsymbol{k}} \propto (k_x,k_y,k_z)$.
	
Here, for the LaIrSi case, we consider both the topological
Hamiltonian and the isotropic spin-triplet pairing.
We focus on the four Fermi surfaces, which are relatively large and
contribute most to the density of states, the latter being highly
relevant to SC. An effective perturbation model around the $R$ point (Figure~\ref{fig:bands}) was constructed,
	%Based on DFT and perturbation, we derive an effective model around the $R$ point, 
whose Hamiltonian is expressed as:
\begin{align}
	\begin{split}
	{\cal H}_{\text{perp}}(\boldsymbol{k}) &= E_0(\boldsymbol{k}) +  ( A_1 k + A_3 |k_xk_yk_z|/k ) s_0\otimes \tau_z  \\
	&+ \lambda_0 \left( \lambda_x s_x + \lambda_y s_y + \lambda_z s_z \right)  \otimes \tau_0, 
	\end{split}
\end{align}
where $E_0=A_0 + A_2 k^2 $, $A_3$ represents the $C_3$-wrapping term, and $\lambda_0$ is the strength of SOC; $s$ and $\tau$ are the Pauli matrices of the spin and the projected orbital degrees of freedom, respectively; $(\lambda_x,\lambda_y,\lambda_z)=\boldsymbol{k}/{k}$ with $\boldsymbol{k}=(k_x,k_y,k_z)$. The energy dispersion are $E_{\alpha,\beta} = E_0  +\alpha (A_1 k +A_3 |k_x k_y k_z|/k) + \beta |\lambda_0|$ with $\alpha, \beta=\pm$. Thus, the $A_3$ term implies that the Fermi surfaces are not isotropic but only symmetric under $C_3$. The band structure of the effective Hamiltonian, shown in blue lines, is presented in Figure~\ref{fig:nodal}a. The three parameters $A_0$, $A_1$, and $A_2$ are extracted from the DFT calculations (red symbols in Figure~\ref{fig:nodal}a). The obtained parameters are $A_0 = 0.536$\,eV, $A_1 = 0.924~$\,eV$\cdot$\AA, $A_2$ = -$14.273~$\,eV$\cdot$\AA$^2$, and $A_3 = 15~$\,eV$\cdot$\AA$^2$
		
	%together with $\lambda_0 = 0$\,eV and $A_3 = 15$\,eV$\cdot$\AA$^2$.
	% The spinless bands with $\lambda_0 = 0$ derived from DFT calculations are depicted in Fig.~\ref{fig:nodal}a (red symbols). The solid lines represent the fits to the bands, by choosing $A_3=15$ eV$\cdot$\AA$^2$.
	
	%\tcr{[JZcomment]: I think the number ``15'' is an arbitary number in a reasonable range, rather from fitting anything.}
	%are depicted in Fig.~\ref{fig:nodal}a for parameter fitting, derived from DFT bands [red dot lines in Fig.~\ref{fig:nodal}a], and we find $A_3=XXX$ eV$\cdot$\AA$^2$. 
	
We then develop an effective Ginzburg-Landau theory for the stabilization of nodal-line SC in LaIrSi, which incorporates both the isotropic pairings ($\Delta_s$ and $\Delta_t$), SOC ($\lambda_0$), and the anisotropic wrapping term ($A_3$) (see details in Note~S5 in the Supporting Information).%As expected, $\Delta_t/\Delta_s \propto \lambda_0$.
With pairing potential $\Delta(\boldsymbol{k})=[\Delta_s + \Delta_t (\boldsymbol{k}\cdot\boldsymbol{s})]is_y\otimes\tau_0$, the superconducting quasi-particle spectrum reads 
\begin{align}\label{eq-main-Ebdg}
	E_{\text{BdG}}^{\alpha,\beta}(\boldsymbol{k})=\pm\sqrt{(E_{\alpha,\beta}(\boldsymbol{k})-E_f)^2+(\Delta_s - \beta \Delta_t k)^2}.
\end{align}
Without loss of generality, here we consider the case with both $\Delta_s > 0$ and $\Delta_t > 0$. The superconducting gap nodes can only appear for the two $\beta = +$ bands, located at the intersection of the surfaces described by the equations $E_{\alpha,+}(\boldsymbol{k}) = E_f$ and $\Delta_s - \Delta_t k =0$. After some
algebra, the condition for nodal-line SC is obtained by imposing $x_{\text{min}}^\pm \le \Delta_s/\Delta_t \le x_{\text{max}}^\pm $ for the $E_{\pm,+}$ band, respectively (see details in Experimental Section and Note~S6, Supporting Information). %The details can be found in the Supplementary materials. 
The evolution of $x_{\text{max}}^-$ and $x_{\text{min}}^-$ versus $\lambda_0$ and their difference versus $A_3$ for the $E_{-,+}$ band are presented in Figure~\ref{fig:nodal}b and Figure~\ref{fig:nodal}c, respectively.
Both $x_{\text{max}}^-$ and $x_{\text{min}}^-$ increase with increasing SOC strength $\lambda_0$. 
	%The red and blue lines represent $x_{\text{max}}^-$ and $x_{\text{min}}^-$, both of which increase with increasing SOC strength $\lambda_0$. 
	%The red line represents $x_{\text{max}}^-$ and and the blue line corresponds to $x_{\text{min}}^-$, and both increase with $\lambda_0$, as depicted in Fig.~\ref{fig:nodal}b.
Notably, the gray region that hosts the gap nodes in Figure~\ref{fig:nodal}b, quantified by the difference between  $x_{\text{max}}^-$ and $x_{\text{min}}^-$, appears to be nearly invariant with respect to $\lambda_0$. 
However, such a difference (i.e., $x_{\text{max}}^-$ -- $x_{\text{min}}^-$)
	%(the gray region) 
increases as the $C_3$-wrapping term $A_3$ increases (Figure~\ref{fig:nodal}c).
	%However, this region undergoes expansion as the $C_3$-wrapping term $A_3$ increases. 
Such conclusions also apply to the $E_{+,+}$ band.
	%These conclusions can be directly applied to the $E_{+,+}$ band.
Moreover, to better illustrate the position of the superconducting nodal lines on the Fermi surfaces, we project the superconducting gaps onto the $E_{\pm,+}$ bands, as depicted in Figure~\ref{fig:nodal}d-f.
For comparison, Figure~\ref{fig:nodal}d shows the two Fermi surfaces corresponding to the $E_{\pm,+}$ bands. 
	%As a comparison, Fig.~\ref{fig:nodal}d corresponds to the two Fermi surfaces of the $E_{\pm,+}$ bands. 
The gap on the $E_{+,+}$ band is nearly isotropic due to the dominant spin-singlet pairing (Figure~\ref{fig:nodal}e), while the gap on the $E_{-,+}$ band exhibits clear nodal lines (Figure~\ref{fig:nodal}f).
Note that, due to the chiral symmetry constraints, there is only
one connected nodal line in each quadrant.
	%
	%\tcr{[Comment]: I've checked the plot code, that the A3=0 and 50 in Fig.4(d), while in Fig.4(e-f), the value is 15. The difference between (e) and (f) is the gap function, with (0.3-kk)*5 and (0.19-kk)*20 respectively. So please comfirm the description above.}
	%in Fig.~\ref{fig:nodal}e is nearly isotropic due to the dominance of spin-singlet pairing, whereas the gap on the $E_{-,+}$ band in Fig.~\ref{fig:nodal}f exhibits nodal-line states. Due to symmetry, there is one connected nodal line in each quadrant. 

Next, we apply the above theoretical framework to the experimental results
of the double-helix La(Rh,Ir)Si chiral materials. 
For convenience, we define $\tilde{\Delta}_t=\Delta_t k_f$.
The temperature-dependent superfluid density $\rho_\mathrm{sc}(T)$ can be well fitted by our theoretical model (Figure~\ref{fig:muSR}e,f), where %$\Delta_s/(\Delta_t k_f) \sim $ 
the dimensionless parameter $\Delta_s/\tilde{\Delta}_t \sim $
1.5 and 0.9
were found to be in  good agreement with the LaRhSi and LaIrSi data, respectively (see details in Experimental Section). 
The analysis of $\rho_\mathrm{sc}(T)$ using different %$\Delta_s/(\Delta_t k_f)$ 
$\Delta_s/\tilde{\Delta}_t$
ratios and other singlet- or triplet-pairing models are reported in Figures~S8 and S9 (Supporting Information).   
	%The fitting to experimental data is discussed in the Method section, where $\Delta_s/\Delta_t \sim \tcr{1.5}$ \tcr{and 0.9 for LaRhSi and LaIrSi, respectively.} \tcr{[LH-comment]: show this value?}
	%Similar features also occur for the other bands \tcr{Shang: Which band?}. 
Our theoretical model agrees well with the experimental observations.
Thus, we conclude that the anisotropic Fermi surfaces with a
nonzero $A_3$ term are crucial for achieving nodal-line SC in
these cubic chiral crystals, in spite of the isotropic
spin-singlet and spin-triplet pairings.
	% are isotropic. 
	
\textbf{Discussion.} We now extend our observations to other superconducting materials with
a chiral space group $P2_13$ (No.~198), thus providing insights for the discovery of new unconventional superconductors.
The reported superconducting materials include binary compounds such as AuBe~\cite{Amon2018,Khasanov2020b}, RhGe~\cite{Mardanya2024,Tsvyashchenko2016}, and ReSi~\cite{Jorda1982}, as well as ternary compounds such as PdBi(Se,Te)~\cite{Lv2019,Joshi2015,Pereti2023}, BaPtP~\cite{Okamoto2020}, and SbPtS~\cite{Mizutani2019}. 
%The binary compounds are isostructural to the CoSi family of chiral crystals, while the ternary ones resemble the La(Rh,Ir)Si chiral crystals, which feature a double-helix structure.
Their electronic band structures and calculation details are presented in Figures~S10--S18 and in Note~S7 (Supporting Information).
The superconducting transition temperatures $T_c$, together with the band-splitting $E_\mathrm{SOC}$ (due to the SOC at the high-symmetry points) are
summarized in Table~S3 (Supporting Information).
%at the high-symmetry $R$ point are summarized in Fig.~\ref{fig:materials}. The values of $E_\mathrm{SOC}$ at the other high-symmetry points and supercondcuting transition temperatures $T_c$s for these chiral crystal materials are provided in Supplementary Table~3. 
According to our DFT calculations, multiple bands cross the Fermi
level in all of these materials. Consequently, they can be classified
as multiband superconductors. 
We have also demonstrated that, in the $\Delta_s/\Delta_t$ vs.\ $\lambda_0$
phase diagram, there are at least two regions where nodal-line SC can be stabilized %for LaIrSi
(see colored regions in Figure~\ref{fig:materials}). These findings may
apply consistently also to other materials. First, LaIrSi and
LaRhSi (indicated by star symbols) are two typical examples
in the phase diagram, characterized by significantly different
band splittings (207 vs.\ 37\,meV) and located in a fully-gapped
and a nodal-line region, respectively. This suggests the possibility of a
crossover from nodal-line SC to fully-gapped SC in the LaIr$_x$Rh$_{1-x}$Si
family,
%It could be interesting to search for the possible crossover from nodal-line SC to fully-gapped SC in LaIr$_x$Rh$_{1-x}$Si family, 
if $\lambda_0$ can be systematically tuned by chemical substitution.  
Second, the binary compound AuBe, with $\lambda_0$ $\sim$ 145\,meV, has been experimentally confirmed to be a fully-gapped superconductor~\cite{Amon2018,Khasanov2020b}. As such, it %Thus, AuBe likely
lies outside the nodal SC regions in the phase diagram. Third, ReSi, despite %having
a large $\lambda_0$ $\sim$ 350\,meV  (comparable to LaIrSi), may
also lie outside the nodal regions, presumably %potentially
positioned in the lower right corner of the phase diagram.
%has a $\lambda_0$ $\sim$ 350\,meV, which is comparable to LaIrSi, it might be located at the lower right corner of the phase diagram, i.e., outside of the nodal regions as well.
Finally, the binary RhGe has a $\lambda_0$ $\sim$ 84\,meV, which is
smaller than that of AuBe and ReSi. Surprisingly, RhGe appears to be located in the first nodal SC region and represents one of the most interesting cases to investigate.

Furthermore, the ternary compounds $TMX$ ($T$ = Co, Ni, Pd, Rh, Ir, and
Pt, $M$ = P, As, Sb, and Bi, $X$ = S, Se, and Te) have the same crystal
structure as La(Rh,Ir)Si~\cite{Hulliger1963}. Their $\lambda_0$s assume
both small and large values, depending on the specific combination of elements.
For instance, PdBiSe ($\lambda_0$ $\sim$ 193\,meV) and PdBiTe ($\lambda_0$ $\sim$ 258\,meV)
become superconducting at $T_c$ = 1.5 and 2.1\,K, respectively~\cite{Joshi2015,Pereti2023}.
In contrast, no physical properties have been reported for PtBiTe ($\lambda_0$ $\sim$ 318\,meV),
belonging to the same family and featuring a very large SOC (see Table~S3, Supporting Information).
Although the $T_c$s are relatively low~\cite{Okamoto2020,Mizutani2019},
SbPtS and BaPtP represent two extreme cases within this family, with
$\lambda_0$ $\sim$ 72 and 321\,meV, respectively. BaPtAs is another interesting case,
with smaller $\lambda_0$ at the $R$ point compared to BaPtP, but similar
$\lambda_0$ values at the $X$ and $\Gamma$ points. While the  cubic
LaIrSi-type phase of BaPtAs does not exhibit SC~\cite{Kudo2018},
its hexagonal form becomes superconducting below $\sim$3\,K. Its
sister compound, BaPtSb, shows the emergence of spontaneous fields,
indicating a breaking of TRS in the SC state~\cite{Sdachi2024}. %Unfortunately,
Except for AuBe and La(Rh,Ir)Si, the superconducting pairing
symmetries of the aforementioned materials remain largely unexplored. 
It is noteworthy that, even in the fully-gapped regions of the phase diagram, unconventional SC can emerge when the SOC strength $\lambda_0$ is significant. 

The phase diagram in Figure~\ref{fig:materials} was constructed using
the material-specific parameters of LaIrSi. Due to variations in
the critical parameters, such as the $A_3$ term, which depends strongly
on the symmetry of the crystal lattice and its electronic structure,
an independent analysis is in high demand for other compounds. 
We emphasize that a combined theoretical and experimental effort, including
spectroscopic and transport probes, is essential to fully resolve
the unconventional superconducting phases across this chiral-crystal
family. %material class.
In conclusion, our results and the summary presented in Figure~\ref{fig:materials} provide a comprehensive roadmap for identifying and exploring unconventional SC in chiral crystals.
%Following our observations, we proposed the locations in the $\Delta_s/\Delta_t$ vs. $\lambda_0$ phase diagram for these superconducting chiral crystals (Fig.~\ref{fig:bands}a), which may guide us to search for novel superconductors with chiral structures. In particular, the XXX and XXX are among the most interesting candidates to show the unconventional nodal SC.  

Due to the structural chirality in real space, chiral crystals also exhibit chirality in momentum space. This leads to
a series of exotic quantum phenomena~\cite{Yan2023,Hasan2021,Narang2021,Bradlyn2016,Chang2018,Tang2017,Lin2022,Fernando2017,Tan2022,Chang2020}.
Yet, due to the limited availability of suitable chiral materials, their unconventional SC and its interplay with the topological electronic bands are barely explored. To date, there are a few theoretical works on the unconventional or topological SC in chiral crystals. 
The binary B20 CoSi family hosts multifold nodes split by the SOC at the same TRIM~\cite{Rao2019,Sanchez2019,Takane2019,Schroter2019,Schroter2020,Yao2020,Sessi2020,Sanchez2023,Krieger2024}. 
%Topological SC arising from the $s_+ \oplus s_-$ pairing has been proposed in RhSi~\cite{Lee2021}. This pairing is $k$-independent, but has opposite signs for the two nodes at the same TRIM. 
In RhGe, the superconducting pairing symmetries involving chiral Fermi pockets indicate their nontrivial characteristics. This makes RhGe a good candidate material for hosting unconventional and topological SC~\cite{Mardanya2024}. Unfortunately, no details of experimental studies have been reported in these B20 chiral crystals. In addition, the structural chirality may also help realize chiral SC too~\cite{Chen2023}. 
The La(Rh,Ir)Si family represents one of the valuable platforms for investigating the chirality-induced topological band topology
and the interplay between weak correlations and geometric quantum effects.
In this family, the correlated phases depend critically on both electron-electron interactions and the quantum geometry/band topology near the Fermi energy, where the chiral structure plays an important role. Different from the RhSi chiral crystal, the La(Rh,Ir)Si family has a double-helix chiral structure with the additional chiral sublattice of the La atoms. This sublattice contributes more electronic states around the Fermi surfaces with a large SOC and triggers the SC. Therefore, the La(Rh,Ir)Si family is one of the few material candidates that can be used to explore the interplay between structural chirality and unconventional SC --- a field of study that combines topological band physics and correlated electron phenomena.
%
%Overall, the unconventional SC are also strongly coupled to the structural chirality induced band topology, and the La(Rh,Ir)Si family represents one of the very rare material candidates to investigate the interplay between structural chirality and unconventional or topological SC--a direction that merges topological band physics with correlated electron phenomena.

%\tcb{Shang: please correct this section.} 
The topological properties of the CoSi family have been extensively studied.
While sharing the same space group with the CoSi family of materials,
La(Rh,Ir)Si exhibit a distinct topological band nature (Figure~\ref{fig:bands}). 
%La(Rh,Ir)Si share the same space group as the CoSi family of materials, their topological band characters exhibit distinct features (see Fig.~\ref{fig:bands}). 
Notably, the effect of SOC is significant in La(Rh,Ir)Si. In the CoSi family, the relatively weak SOC obscures the splitting between the spin-3/2 multifold and the Weyl node, making it challenging to resolve them via photoemission spectroscopy~\cite{Rao2019,Sanchez2019,Takane2019}. 
The pronounced SOC, on the other hand, causes a significant
splitting in LaIrSi, which is likely to be detectable using advanced photoemission techniques. 
%which may be detectable using advanced photoemission technique. 
Further distinctions emerge in the electronic excitations. The spin-1 excitation with sixfold degeneracy is located at $\sim 138$\,meV above the Fermi level in La(Rh,Ir)Si. 
Additionally, a fourfold degeneracy lying just $\sim$42~meV below
the Fermi level and with a topological charge $C = +2$ resides at the
zone center, but has never been observed in the CoSi family.
These unconventional topological features, including the proximity of these states to the Fermi level, may serve as the driving forces behind the emergence of intrinsic topological superconductors and the long-sought Majorana quasiparticles.	
Finally, it would be interesting to investigate the chirality-related superconducting phenomena in LaIrSi-based devices. 
	A prime example is the Josephson diode effect that requires simultaneous breaking of inversion and mirror symmetries.
	The La(Rh,Ir)Si family  offers a natural platform to realize it. For example, TaS$_2$ intercalated with chiral molecules shows a distinct superconducting diode effect, while it is absent in the pristine achiral TaS$_2$~\cite{Wan2024}.

\section{Conclusion}

In summary, we observed unconventional superconductivity in the double-helix La(Rh,Ir)Si chiral crystals, which exhibit exotic multifold fermions near the Fermi level and giant helicoid Fermi arcs that span the entire surface Brillouin zone. %The presence of bulk SC is closely related to the additional contributions of the La-$5d$ orbitals compared to the CoSi family of chiral crystals.    
%particularly in LaRhSi (fully gapped) and LaIrSi (nodal-line), 
Our work reveals two dominant factors for the emergence of novel superconducting phases in La(Rh,Ir)Si, distinct from the conventional superconducting behavior observed in AuBe, despite sharing the same chiral space group. 
The first factor is the double-helix chiral structure.
This enhances the density of electronic states near the
Fermi level due to the additional contributions of the
La-$5d$ orbitals compared to the CoSi family of chiral crystals. 
The second factor involves tuning the ratio of spin-singlet to spin-triplet pairing via the spin-orbit coupling.
This causes superconducitivity to evolve from a fully-gapped state (in LaRhSi) to a nodal-line state (in LaIrSi).
Moreover, we develop a phenomenological theory that elucidates a new mechanism
for nodal-line SC (here arising from the presence of anisotropic Fermi surfaces)  
and reproduces the experimental data remarkably well.
Our work opens up a new avenue for exploring unconventional or
topological superconductivity in chiral materials. At the same time,
it calls for a re-examination of the interplay between SC and topologically non-trivial Fermi surfaces, an interplay that has hardly been considered in other noncentrosymmetric superconductors with a chiral crystal structure.
%\tcr{Our complementary imaging and electrical transport studies
%	provide clear evidence for the correlation between topological Hall resistivity and the
%	topology of the magnetic texture in technologically viable magnetic
%	multilayers.}

% Experimental section

\section{Experimental Section}
\noindent\threesubsection{Materials synthesis and characterization}\\
Polycrystalline  LaRhSi and LaIrSi samples were prepared by
stoichiometric arc melting of a La rod (99.9\%, Alfa Aesar), Rh or Ir powders (99.9\%, ChemPUR), and Si chunks (99.9999\%, Alfa Aesar) in a high-purity argon atmosphere. To improve sample homogeneity, the ingots were flipped and re-melted more than six times. 
The crystal structure of the LaRhSi and LaIrSi samples was checked via powder 
x-ray diffraction at room temperature using a Bruker D8 diffractometer 
with Cu K$\alpha$ radiation. This confirmed the cubic
non\-cen\-tro\-sym\-met\-ric structure of LaRhSi and LaIrSi ($P$2$_1$3, No.\,198),
which is also of chiral type. The electrical-resistivity and magnetic susceptibility measurements were performed on a Quantum Design physical property measurement system (PPMS) and a magnetic property measurement system (MPMS).\\

\noindent\threesubsection{{\textmu}SR experiments}\\
The {\textmu}SR measurements were performed at the general-purpose surface-muon spectrometer (GPS) at the $\pi$M3 beam line and the multipurpose surface-muon spectrometer (Dolly) at the $\pi$E1 beamline of the Swiss muon source (S{\textmu}S) at Paul Scherrer Institut (PSI) in Villigen, Switzerland. 
To exclude the possibility of stray magnetic fields during the
ZF-{\textmu}SR measurements, all the magnets were preliminarily degaussed,
and we made use of an active field-compensating facility~\cite{Amato2017}. 
For the TF-{\textmu}SR measurements, the applied magnetic field (i.e., 20\,mT) was perpendicular to the muon-spin direction and the samples were cooled in an applied magnetic field down to the base temperature ($\sim$0.3\,K). Both the ZF- and TF-{\textmu}SR spectra were collected upon heating the samples.\\

\noindent\threesubsection{Electronic band-structure calculations}\\
First-principles calculations were performed based on the density functional
theory (DFT), as implemented in the Vienna ab-initio simulation package
(VASP) package~\cite{Kresse1996kl,Kresse1996vk}. Projector
augmented wave (PAW) pseudo-potentials were adopted in the calculation~\cite{Kresse1999wc,Blochl1994zz}.
The generalized gradient approximation with the Perdew-Burke-Ernzerhof (PBE) realization~\cite{Perdew1996iq} was used for the exchange-correlation functional.
The valence electrons treated in the calculations include La ($5s^26s^25p^65d^1$), Rh ($4d^85s^1$), Ir ($5d^86s^1$), and Si ($3s^23p^2$). The kinetic energy
cutoff was fixed to 400 eV. For the self-consistent calculations, the
Brillouin zone (BZ) integration was performed on a $\Gamma$-centered mesh
of $12 \times 12 \times 12$ $k$-points. The energy convergence criteria
was set to 10$^{-7}$\,eV.
In order to study the topological electronic structure of LaIrSi, a Wannier tight binding Hamiltonian consisting of La-$5d$, La-$6s$, Ir-$5d$, Si-$3s$ and Si-$3p$ orbitals was constructed using the Wannier90 package~\cite{pizzi_wannier90_2020}.
The surface spectra (i.e., Fermi arcs) were calculated by using the iterative Green's function method as implemented in the WannierTools package~\cite{wu_wanniertools_2018}. \\

\noindent\threesubsection{Material phase diagram}\\
In Figure~\ref{fig:materials}, we qualitatively map out the phase
diagram of unconventional SC in both simple- %chiral
and in double-helix chiral crystals. Our analysis primarily focuses
on identifying the regions where nodal-line SC emerges. To achieve this,
we rely on the spectrum of the BdG Hamiltonian, as described by Equation~\ref{eq-main-Ebdg}.
For the case with both $\Delta_s$ $>$ 0 and $\Delta_t$ $>$ 0, the superconducting nodes can only occur for the two $\beta = +$ bands, and the location of the nodes is % the location is ...
determined by solving 
\begin{align}
	\begin{cases}
		E_{\alpha,+}(\boldsymbol{k}) = E_f, \\
		\Delta_s - \Delta_t k =0.
	\end{cases}
\end{align}
More explicitly, these conditions can be expressed as
\begin{align}
	\begin{cases}
		\alpha |k_xk_yk_z| = k(\tilde{E_f} - \alpha \tilde{A}_1k + \tilde{A}_2k^2), \\ 	k = \frac{\Delta_s}{\Delta_t},
	\end{cases}
\end{align}
where we define $\tilde{E_f}=( E_f - A_0 - |\lambda_0|)/A_3<0$, $\tilde{A}_1= A_1/A_3>0$, and $\tilde{A}_2 = -A_2/A_3>0$. Because of the three glide symmetries about the $x$--$z$, $y$--$z$ and $x$--$y$ planes, it is sufficient to analyze the quadrant where $k_x>0, k_y>0, k_z>0$. 

\begin{figure}
	  \centering 
	\includegraphics[width=0.8\linewidth]{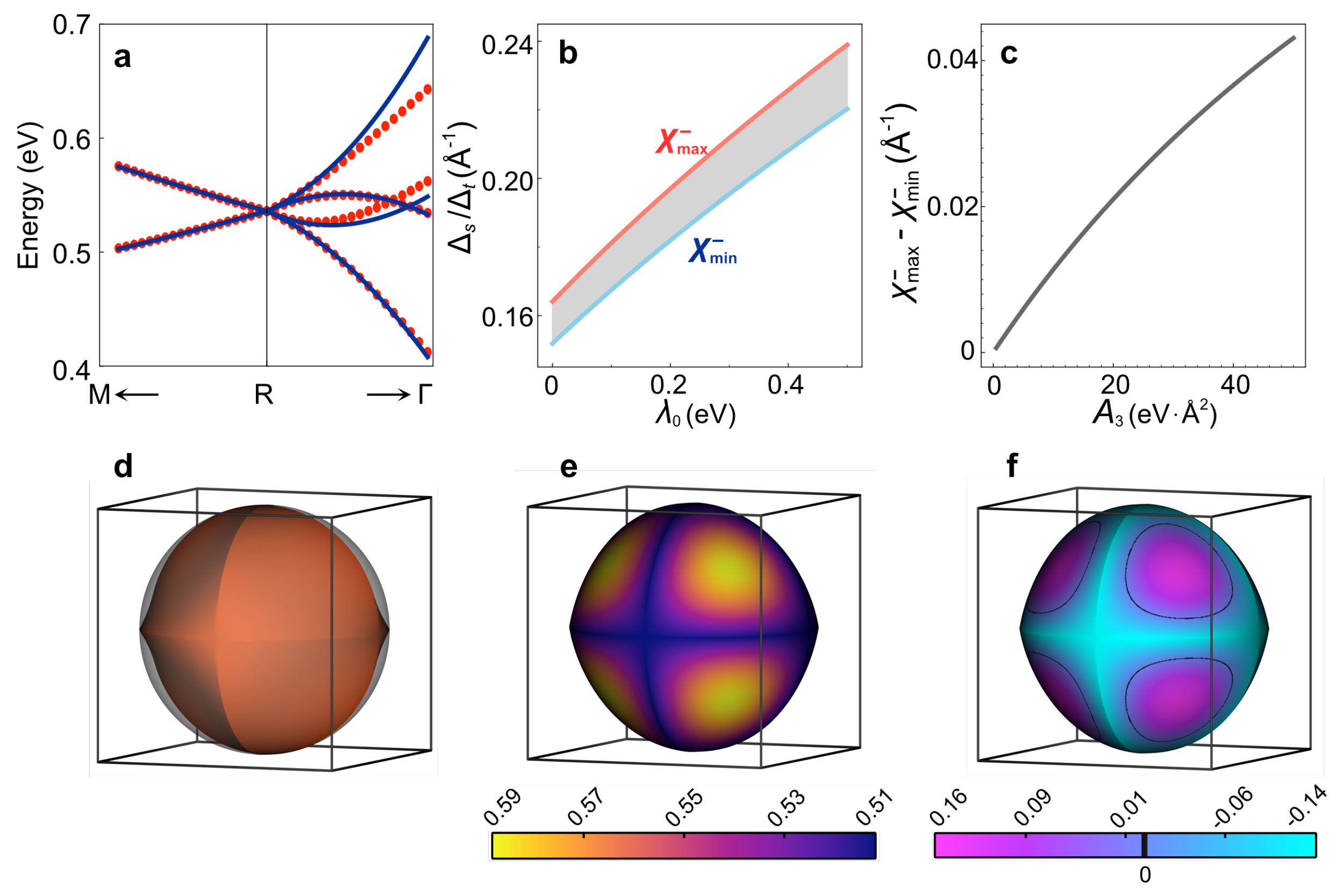}
	\caption{\label{fig:nodal} Calculations of  nodal-line superconductivity. a) Band structure of the effective Hamiltonian around the $R$ point (blue lines), and the parameters fitted from the DFT calculated bands (red symbols). 
		b) The superconducting phase diagram for the $E_{-,+}$ band as a function of $\Delta_s/\Delta_t$ and $\lambda_0$. Nodal-line superconductivity (gray region) emerges when $\Delta_s/\Delta_t$ lies in the region between the minimum $x^{-}_\mathrm{min}$ (blue line) and maximum $x^{-}_\mathrm{max}$ (red line).
		c) The difference between $x^{-}_\mathrm{max}$ and $x^{-}_\mathrm{min}$ as a function of the parameter $A_3$. 
		d) Comparison plot of the isotropic (gray sphere) and anisotropic (orange sphere) Fermi surfaces. To better illustrate the
		anisotropy, here we use a large $A_3 = 50$\,eV$\cdot$\AA$^2$ parameter. e) Gap function of the fully-gapped superconducting anisotropic Fermi surface. f) Gap function of the nodal superconducting anisotropic Fermi surface. The nodal lines are shown as black lines. The color bars show the energy in meV.}
\end{figure}

\begin{figure}
		  \centering 
	\includegraphics[width=0.7\linewidth]{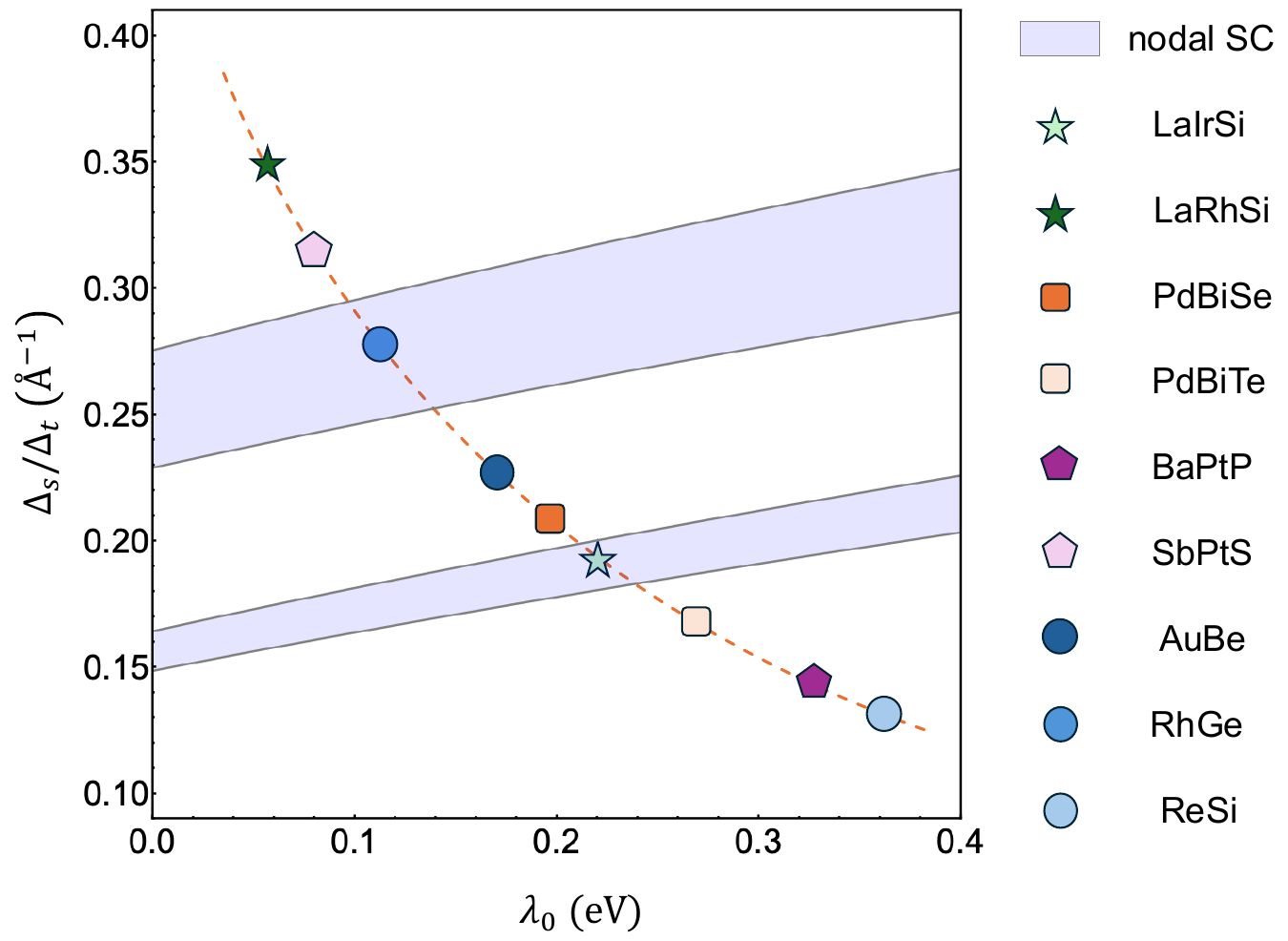}
	\caption{\label{fig:materials} Proposed $\Delta_s/\Delta_t$ vs. $\lambda_0$ phase diagram. Superconducting materials with a standard chiral (e.g., AuBe) and a double-helix chiral (e.g., LaIrSi) structure depicted %summarized
		in the $\Delta_s/\Delta_t$ vs. $\lambda_0$ phase diagram.  
		The latter may act as a guide in the search for unconventional SC in chiral crystals. Here, $\lambda_0$ is the band spin-splitting strength at the high-symmetry $R$ point. The colored regions indicate the nodal bands, corresponding to $\Delta_s$/$\Delta_t$ $\in$ [$x_{\text{min}}^{-}$, $x_{\text{max}}^{-}$] and [$x_{\text{min}}^{+}$, $x_{\text{max}}^{+}$] (see details in Figure~\ref{fig:nodal}b).
		The superconducting ground state of these materials has been experimentally confirmed. 
		The ratio $\Delta_s/\Delta_t$ is computed from the effective model for La(Rh,Ir)Si. 
		The detailed calculations are presented in the Experimental Section and Note S6 in the Supporting Information. The dashed line is a guide for the eyes.}
\end{figure}

After performing straightforward calculations (whose details are
provided in Note S6, Supporting Information), we derive the conditions for nodal-line SC:
\begin{itemize}
	\item For the $E_{+,+}$ band (where $\alpha=+$). The condition for nodal-line SC is given by
	\begin{align}
		x_\mathrm{min}^+ \le \frac{\Delta_s}{\Delta_t}	 \le  x_\mathrm{max}^+,
	\end{align}
	where 
	\begin{align}
		x_\mathrm{min} ^+&= \frac{1}{2\tilde{A}_2}\left(\tilde{A}_1 + \sqrt{\tilde{A}_1^2 - 4\tilde{A}_2\tilde{E}_f} \right), \\
		x_\mathrm{max}^+ &= \frac{1}{2 \tilde{A}_2'}
		\left( \tilde{A}_1 + \sqrt{ \tilde{A}_1^2 - 4\tilde{E}_f \tilde{A}_2' } \right).
	\end{align}
	Here, we define $\tilde{A}_2' = \tilde{A}_2-\frac{\sqrt{3}}{9}$. In this work, we focus on the case with $\tilde{A}_2'>0$. 
	\item For the $E_{-,+}$ band ($\alpha=-$). Likewise, the condition for nodal-line SC is
	\begin{align}
		x_\mathrm{min}^- \le \frac{\Delta_s}{\Delta_t}	 \le  x_\mathrm{max}^-,
	\end{align}
	where 
	\begin{align}
		x_\mathrm{min} ^- &= \frac{1}{2 \tilde{A}_2''}
		\left( -\tilde{A}_1 + \sqrt{ \tilde{A}_1^2 - 4\tilde{E}_f \tilde{A}_2'' } \right), \\
		x_\mathrm{max}^- &=\frac{1}{2\tilde{A}_2}\left(-\tilde{A}_1 + \sqrt{\tilde{A}_1^2 - 4\tilde{A}_2\tilde{E}_f} \right).
	\end{align}
	Here, we define $\tilde{A}_2'' = \tilde{A}_2+\frac{\sqrt{3}}{9}$. The dependence of $x_\mathrm{min}^-$ and $x_\mathrm{max}^-$ on $\lambda_0$ and $A_3$ is % the dependence is - singular
	presented in Figure~\ref{fig:nodal}. Notably, as discussed in the main text,
	the nodal region expands as the $C_3$-wrapping term $A_3$ increases.
	The same features are found for the $E_{+,+}$ band ($\alpha = +$) (see Figure~\ref{fig:materials}).
\end{itemize}

In the derivation of the nodal-line superconductivity scenario above, we treat $\Delta_s/\Delta_t$ as an independent tuning parameter, which can be determined either through self-consistent calculations or by fitting experimental data. Here, we present a general comparison between LaRhSi and LaIrSi. When $\lambda_0\to0$, the ratio $\Delta_s/\Delta_t$ is relatively large (i.e., $\Delta_s/\Delta_t \gg x_\text{max}^{\pm}$), which explains the fully-gapped superconducting state observed in LaRhSi (a system with weak SOC). As $\lambda_0$ increases, we expect that the ratio $\Delta_s/\Delta_t$ decreases, leading to the emergence of nodal SC in the $E_{+,-}$ band. Furthermore, our calculations reveal that the $E_{-,-}$ band can also exhibit nodal SC when $\lambda_0$ increases further. This suggests that the relatively strong SOC in LaIrSi reduces $\Delta_s/\Delta_t$, thereby enabling the possibility of nodal SC.\\

\noindent\threesubsection{Approximation to fit the superfluid density}\\
Experimentally, the ratio $\Delta_s/\Delta_t$ can be estimated 
directly from the measured superfluid density data. In the previous section,
the ratio $\Delta_s/\Delta_t$ was expressed in units of [{\AA$^{-1}$}],
since the dimensions of both $[\Delta_s]$ and $[\Delta_t\times k_f]$ are [eV], where $k_f$ is the Fermi momentum. This formulation is more general for analyzing the phase diagram (Figure~\ref{fig:materials}), as it accounts for both the inner and outer Fermi surfaces. However, to fit the experimental data, a dimensionless parameter 
%$f_\mathrm{st}=\Delta_s/(\Delta_tk_f)$ 
$\Delta_s/\tilde{\Delta}_t$ with $\tilde{\Delta}_t=\Delta_t k_f$
is more convenient, as will be illustrated below.

In addition, the presence of an anisotropic band structure or Fermi surfaces
complicates the analysis, making a simple fitting approach impractical.
To address this (i.e., $k_f$ is not isotropic), we devise
an approximation by mapping the non-spherical Fermi surface onto a spherical
Fermi surface. While this approximation simplifies the geometry, it introduces an anisotropic gap function. First, we solve the equation 
\begin{align}
	E_0  + \alpha (A_1 k +A_3 |k_x k_y k_z|/k) + |\lambda_0|  = E_f,
\end{align}
for the Fermi momentum $k$, which leads to the expression
\begin{align}\label{eq-method-kthetaphi}
	k(\theta,\phi) = -\frac{2 E_f^0}{-\alpha A_1 + \sqrt{ A_1^2 + 4 A_2(\theta,\phi) E_f^0 }},
\end{align}
where we define $E_f^0 = E_f - A_0 - |\lambda_0| $ and $A_2(\theta,\phi)=A_2+\alpha \frac{A_3}{4}|\sin 2\theta \sin 2\phi \sin \theta|$. Note that, $A_0<0$, $A_1>0$, $A_2<0$ and $A_3>0$, so that $E_f^0<0$ and $A_2(\theta,\phi) \le A_2 + \frac{\sqrt{3}}{9} A_3  <0$.

Then, we figure out the pairing gap projected onto the Fermi surfaces. For the intra-band pairing gap, we have
\begin{align} \label{eq-method-delta}
	\Delta(\theta,\phi) = \Delta_s - \tilde{\Delta}_t \frac{k_{\text{min}}}{k(\theta,\phi)},
\end{align}
where we define $k_{\text{min}}$ as the smallest Fermi momentum of $k(\theta,\phi)$
for $\theta\in [0, \pi)$  %]
and $\phi\in   [0,2\pi)$. %]
%Note that $\Delta_s/\Delta_t$ in this context represents the dimensionless parameter $f_\mathrm{st}$. 
The value of $k_{\text{min}}$ varies across different bands. For the $E_{+,-}$ band (where $\alpha = +$), it is given by
\begin{align}\label{eq-method-kmin-1}
	k_{\text{min}} = -\frac{2 E_f^0}{-A_1 + \sqrt{A_1^2 + 4A_2 E_f^0}}.
\end{align}
Finally, by substituting Equation~\ref{eq-method-kthetaphi} and Equation~\ref{eq-method-kmin-1} into Equation~\ref{eq-method-delta}, we obtain the fitting gap function for the $E_{+,-}$ band as
\begin{align}\label{eq-gapfun1}
	\Delta(\theta,\phi) = \Delta_s - \tilde{\Delta}_t \frac{ A_1 - \sqrt{A_1^2 + 4A_2(\theta,\phi) E_f^0} }{A_1 - \sqrt{A_1^2 + 4A_2 E_f^0}},
\end{align}
where $A_2(\theta,\phi)  = A_2 + \frac{A_3}{4}|\sin 2\theta \sin 2\phi \sin \theta|$.\\
Likewise, for the $E_{-,-}$ band ($\alpha = -$), $k_{\text{min}}$ is given by
\begin{align}
	k_{\text{min}} = -\frac{2 E_f^0}{A_1 + \sqrt{A_1^2 + 4(A_2-\frac{A_3}{3\sqrt{3}}) E_f^0}},
\end{align}
because $|\sin 2\theta \sin 2\phi \sin \theta|_\mathrm{max} =  4/(3 \sqrt{3})$. Then, the fitting gap function for the $E_{-,-}$ band is 
\begin{align} \label{eq-gapfun2}
	\Delta(\theta,\phi) = \Delta_s - \tilde{\Delta}_t \frac{ A_1 + \sqrt{A_1^2 + 4A_2(\theta,\phi) E_f^0} }{A_1 + \sqrt{A_1^2 + 4(A_2-\frac{A_3}{3\sqrt{3}}) E_f^0}},
\end{align}
where $A_2(\theta,\phi)  = A_2 - \frac{A_3}{4}|\sin 2\theta \sin 2\phi \sin \theta|$.

\begin{figure}
	  \centering 
	\includegraphics[width=0.90\linewidth]{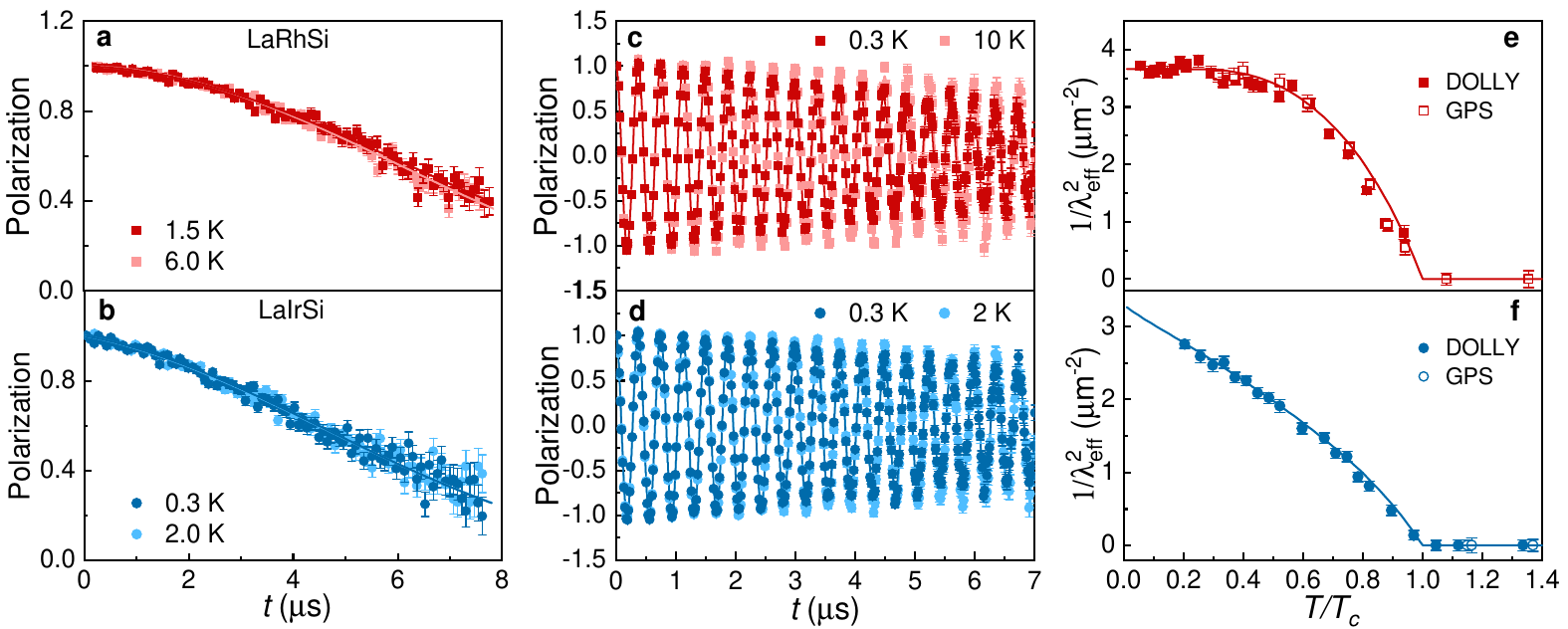}
	\caption{\label{fig:muSR} ZF-{\textmu}SR, TF-{\textmu}SR and superfluid density. 
		 a,b) Zero-field {\textmu}SR spectra collected above and below $T_c$ for LaRhSi (a) and LaIrSi (b). The almost overlapping spectra, indicate the absence of an additional muon-spin relaxation and, thus, the absence of TRS breaking in the superconducting state. Solid lines through the data are fits to Equation~1 in Note~S2 in the Supporting Information. 
		 c,d) Transverse-field {\textmu}SR spectra collected in the  superconducting and normal states (i.e., above and below $T_c$) of LaRhSi (c) and LaIrSi (d). A magnetic field of 20\,mT was applied in the normal state, and the TF-{\textmu}SR spectra were collected upon heating the sample. The enhanced muon-spin relaxation rate in the superconducting state is due to the formation of FLL. Solid lines through the data are fits to Equation~2 in Note~S2 in the Supporting Information.
		 e,f) Superfluid density [$\rho_\mathrm{sc}(T) \propto \lambda_\mathrm{eff}^{-2}(T)$] as a function of reduced temperature $T/T_c$ for LaRhSi (e) and LaIrSi (f). Data acquired at the Dolly and GPS spectrometers are highly consistent. Solid lines represent fits to Equation~3 in Note~S2 in the Supporting Information with a singlet- and triplet-dominated pairing for LaRhSi and LaIrSi, respectively, with the gap function shown in the Experimental Section.
		The derived fitting parameters for both compounds are listed in Table~S2 in the Supporting Information. The error bars of  $\lambda_\mathrm{eff}^{-2}(T)$ are the SDs obtained from fits of the TF-{\textmu}SR spectra by the \texttt{musrfit} software package~\cite{Suter2012}.   
	} 
\end{figure}

The key parameters obtained from the fit are $E_f = 0$, $A_0 = 0.536 \; \text{eV}$, $A_1 =  0.924 \; \text{eV}$, $A_2 = -14.273 \; \text{eV}\cdot$\AA$^2$, $\lambda_0|_{\text{LaRhSi}} = 0.1 \; \text{eV}$, $\lambda_0|_{\text{LaIrSi}} = 0.4 \; \text{eV}$, and $A_3 = 15 \; \text{eV}\cdot$\AA$^2$. These best-fit results for $\Delta_s/\tilde{\Delta}_t$ %\tcr{using Eq.~\eqref{eq-gapfun2}} 
are presented in the main text (see Figure~\ref{fig:muSR}). %\tcr{Note that, using Eq.~\eqref{eq-gapfun1} leads comparable results.}
Our model nicely captures the evolution of fully-gapped to nodal-line SC in the La(Rh,Ir)Si family.\\

%\medskip
%\textbf{Supporting Information} \par %Please delete the Suppporting Information statement if it is not applicable. Please supply Supporting Information in another file. Supporting information should not be provided in .tex format
%Supporting Information is available from the Wiley Online Library or from the author.

% Acknowledgements
%\medskip
%\textbf{Acknowledgements} \par %delete if not applicable))
%This work was supported by the Natural Science Foundation of Shanghai (Grants No.21ZR1420500 and 21JC1402300), Natural Science Foundation of Chongqing (Grant No.2022NSCQ-MSX1468), the National Natural Science Foundation of China (No. 12174103). Y.X. acknowledges support from the Shanghai Pujiang Program (Grant No.21PJ1403100).

\medskip
\noindent\textbf{Conflict of Interest}\\
The authors declare no conflict of interest.

\medskip
\noindent\textbf{Data Availability Statement} \\
The data that support the findings of this study are available from the corresponding author upon reasonable request.

\medskip
\noindent\textbf{Supporting Information}\\% \par %Please delete the Suppporting Information statement if it is not applicable. Please supply Supporting Information in another file. Supporting information should not be provided in .tex format
Supporting Information is available from the Wiley Online Library or from the author.

% Acknowledgements
\medskip
\noindent\textbf{Acknowledgements}\\%\par %delete if not applicable))
We acknowledge the allocation of beam time at S$\mu$S (Dolly and GPS
spectrometers) and thank Sudeep K. Ghosh for useful discussions.
T.S.\ acknowledges support by the National Natural Science Foundation of China
(Grant Nos.\ 12374105 and 12350710785),
the Natural Science Foundation of Shanghai (Grant Nos.\ 21ZR1420500 and
21JC1402300), the Natural Science Foundation of Chongqing (Grant No.\
2022NSCQ-MSX1468), and the Fundamental Research Funds for the Central
Universities.
%and the Schweizerische Nationalfonds zur F\"{o}r\-der\-ung der
%Wis\-sen\-schaft\-lichen For\-schung (SNF) (Grants No.\ 200021\_169455 and No.\ 200021\_188706). We acknowledge the allocation of beam time at the Swiss muon source (GPS {\textmu}SR spectrometer).
\medskip
% References
% Use the following code if you wish to generate your bibliography with BibTeX;
% replace the string "MSP-template" below with the name(s) of
% the BibTeX data base(s) you want to use.
% The resulting bibliography-output (the content of the .bbl file)
% must be pasted back into this file before submission.
% Please also include your BibTeX data base file(s) in your submission
% so that we can re-run BibTeX if necessary.
%
\bibliographystyle{MSP}

\bibliography{LaIrSi.bib}

\begin{thebibliography}{10}
\providecommand{\url}[1]{\texttt{#1}}
\providecommand{\urlprefix}{URL }

\bibitem{Siegel1998}
J.~S. Siegel,
\newblock \emph{Chirality} \textbf{1998}, \emph{10} 24.

\bibitem{Bradlyn2016}
B.~Bradlyn, J.~Cano, Z.~Wang, M.~G. Vergniory, C.~Felser, R.~J. Cava, B.~A.
  Bernevig,
\newblock \emph{Science} \textbf{2016}, \emph{353} aaf5037.

\bibitem{Chang2018}
G.~Chang, B.~J. Wieder, F.~Schindler, D.~S. Sanchez, I.~Belopolski, S.-M.
  Huang, B.~Singh, D.~Wu, T.-R. Chang, T.~Neupert, S.-Y. Xu, H.~Lin, M.~Z.
  Hasan,
\newblock \emph{Nat. Mater.} \textbf{2018}, \emph{17} 978.

\bibitem{Fecher2022}
G.~H. Fecher, J.~K\"{u}bler, C.~Felser,
\newblock \emph{Materials} \textbf{2022}, \emph{15} 5812.

\bibitem{Naaman2019}
R.~Naaman, Y.~Paltiel, D.~H. Waldeck,
\newblock \emph{Nat. Rev. Chem.} \textbf{2019}, \emph{3} 250.

\bibitem{Wang2023}
X.~Wang, C.~Yi, C.~Felser,
\newblock \emph{Adv. Mater.} \textbf{2023}, 2308746.

\bibitem{Dryzun2012}
C.~Dryzun, D.~Avnir,
\newblock \emph{Chem. Commun.} \textbf{2012}, \emph{48} 5874.

\bibitem{Hasan2021}
M.~Z. Hasan, G.~Chang, I.~Belopolski, G.~Bian, S.-Y. Xu, J.-X. Yin,
\newblock \emph{Nat. Rev. Mater.} \textbf{2021}, \emph{6} 784.

\bibitem{Narang2021}
P.~Narang, C.~A.~C. Garcia, C.~Felser,
\newblock \emph{Nat. Mater.} \textbf{2021}, \emph{20} 293.

\bibitem{Kumar2021}
N.~Kumar, S.~N. Guin, K.~Manna, C.~Shekhar, C.~Felser,
\newblock \emph{Chem. Rev.} \textbf{2021}, \emph{121} 2780.

\bibitem{Yan2023}
B.~Yan,
\newblock Structural chirality and electronic chirality in quantum materials,
  \textbf{2023},
\newblock ArXiv:2312.03902.

\bibitem{Yang2024}
Q.~Yang, Y.~Li, C.~Felser, B.~Yan,
\newblock Chirality induced spin selectivity in chiral crystals, \textbf{2024},
\newblock ArXiv:2312.04366.

\bibitem{Yan2017}
B.~Yan, C.~Felser,
\newblock \emph{Annu. Rev. Condens. Matter Phys.} \textbf{2017}, \emph{8} 337.

\bibitem{Armitage2018}
N.~P. Armitage, E.~J. Mele, A.~Vishwanath,
\newblock \emph{Rev. Mod. Phys.} \textbf{2018}, \emph{90} 015001.

\bibitem{Nagaosa2020}
N.~Nagaosa, T.~Morimoto, Y.~Tokura,
\newblock \emph{Nat. Rev. Mater.} \textbf{2020}, \emph{5} 621.

\bibitem{Lv2021}
B.~Q. Lv, T.~Qian, H.~Ding,
\newblock \emph{Rev. Mod. Phys.} \textbf{2021}, \emph{93} 025002.

\bibitem{Amato1997}
A.~Amato,
\newblock \emph{Rev. Mod. Phys.} \textbf{1997}, \emph{69} 1119.

\bibitem{Nagaosa2013}
N.~Nagaosa, Y.~Tokura,
\newblock \emph{Nat. Nanotechnol.} \textbf{2013}, \emph{8} 899.

\bibitem{Fert2017}
A.~Fert, N.~Reyren, V.~Cros,
\newblock \emph{Nat. Rev. Mater.} \textbf{2017}, \emph{2} 17031.

\bibitem{Yang2021}
S.-H. Yang, R.~Naaman, Y.~Paltiel, S.~S.~P. Parkin,
\newblock \emph{Nat. Rev. Phys.} \textbf{2021}, \emph{3} 328.

\bibitem{Tokura2021}
Y.~Tokura, N.~Kanazawa,
\newblock \emph{Chem. Rev.} \textbf{2021}, \emph{121} 2857.

\bibitem{Cheong2022}
S.-W. Cheong, X.~Xu,
\newblock \emph{npj Quantum Mater.} \textbf{2022}, \emph{7} 40.

\bibitem{Zhang2025}
K.-X. Zhang, S.~Cheon, H.~Kim, P.~Park, Y.~An, S.~Son, J.~Cui, J.~Keum,
  J.~Choi, Y.~Jo, H.~Ju, J.-S. Lee, Y.~Lee, M.~Avdeev, A.~Kleibert, H.-W. Lee,
  J.-G. Park,
\newblock \emph{Phys. Rev. Lett.} \textbf{2025}, \emph{134} 176701.

\bibitem{An2023}
Y.~An, P.~Park, C.~Kim, K.~Zhang, H.~Kim, M.~Avdeev, J.~Kim, M.-J. Han, H.-J.
  Noh, S.~Seong, J.-S. Kang, H.-D. Kim, J.-G. Park,
\newblock \emph{Phys. Rev. B} \textbf{2023}, \emph{108} 054418.

\bibitem{Kim2024}
J.~Kim, K.~Zhang, P.~Park, W.~Cho, H.~Kim, J.-G. Park,
\newblock Electrical control of topological {3Q} state in an intercalated van
  der {Waals} antiferromagnet, \textbf{2024},
\newblock ArXiv:2409.02710.

\bibitem{Tang2017}
P.~Tang, Q.~Zhou, S.-C. Zhang,
\newblock \emph{Phys. Rev. Lett.} \textbf{2017}, \emph{119} 206402.

\bibitem{Lin2022}
M.~Lin, I.~n. Robredo, N.~B.~M. Schr\"oter, C.~Felser, M.~G. Vergniory,
  B.~Bradlyn,
\newblock \emph{Phys. Rev. B} \textbf{2022}, \emph{106} 245101.

\bibitem{Fernando2017}
F.~De~Juan, A.~G. Grushin, T.~Morimoto, J.~E. Moore,
\newblock \emph{Nat. Commun.} \textbf{2017}, \emph{8} 15995.

\bibitem{Tan2022}
W.~Tan, X.~Jiang, Y.~Li, X.~Wu, J.~Wang, B.~Huang,
\newblock \emph{Adv.Fun. Mater.} \textbf{2022}, \emph{32} 2208023.

\bibitem{Chang2020}
G.~Chang, J.-X. Yin, T.~Neupert, D.~S. Sanchez, I.~Belopolski, S.~S. Zhang,
  T.~A. Cochran, Z.~c. v. b.~a. Ch\'eng, M.-C. Hsu, S.-M. Huang, B.~Lian, S.-Y.
  Xu, H.~Lin, M.~Z. Hasan,
\newblock \emph{Phys. Rev. Lett.} \textbf{2020}, \emph{124} 166404.

\bibitem{Rao2019}
Z.~Rao, H.~Li, T.~Zhang, S.~Tian, C.~Li, B.~Fu, C.~Tang, L.~Wang, Z.~Li,
  W.~Fan, J.~Li, Y.~Huang, Z.~Liu, Y.~Long, C.~Fang, H.~Weng, Y.~Shi, H.~Lei,
  Y.~Sun, T.~Qian, H.~Ding,
\newblock \emph{Nature} \textbf{2019}, \emph{567} 496.

\bibitem{Sanchez2019}
D.~S. Sanchez, I.~Belopolski, T.~A. Cochran, X.~Xu, J.-X. Yin, G.~Chang,
  W.~Xie, K.~Manna, V.~S\"{u}\ss, C.-Y. Huang, N.~Alidoust, D.~Multer, S.~S.
  Zhang, N.~Shumiya, X.~Wang, G.-Q. Wang, T.-R. Chang, C.~Felser, S.-Y. Xu,
  S.~Jia, H.~Lin, M.~Z. Hasan,
\newblock \emph{Nature} \textbf{2019}, \emph{567} 500.

\bibitem{Takane2019}
D.~Takane, Z.~Wang, S.~Souma, K.~Nakayama, T.~Nakamura, H.~Oinuma, Y.~Nakata,
  H.~Iwasawa, C.~Cacho, T.~Kim, K.~Horiba, H.~Kumigashira, T.~Takahashi,
  Y.~Ando, T.~Sato,
\newblock \emph{Phys. Rev. Lett.} \textbf{2019}, \emph{122} 076402.

\bibitem{Sanchez2023}
D.~S. Sanchez, T.~A. Cochran, I.~Belopolski, Z.-J. Cheng, X.~P. Yang, Y.~Liu,
  T.~Hou, X.~Xu, K.~Manna, C.~Shekhar, J.-X. Yin, H.~Borrmann, A.~Chikina,
  J.~D. Denlinger, V.~N. Strocov, W.~Xie, C.~Felser, S.~Jia, G.~Chang, M.~Z.
  Hasan,
\newblock \emph{Nat. Phys.} \textbf{2023}, \emph{19} 682.

\bibitem{Schroter2019}
N.~B.~M. Schr\"{o}ter, D.~Pei, M.~G. Vergniory, Y.~Sun, K.~Manna, F.~De~Juan,
  J.~A. Krieger, V.~S\"{u}ss, M.~Schmidt, P.~Dudin, B.~Bradlyn, T.~K. Kim,
  T.~Schmitt, C.~Cacho, C.~Felser, V.~N. Strocov, Y.~Chen,
\newblock \emph{Nat. Phys.} \textbf{2019}, \emph{15} 759.

\bibitem{Schroter2020}
N.~B.~M. Schr\"{o}ter, S.~Stolz, K.~Manna, F.~De~Juan, M.~G. Vergniory, J.~A.
  Krieger, D.~Pei, T.~Schmitt, P.~Dudin, T.~K. Kim, C.~Cacho, B.~Bradlyn,
  H.~Borrmann, M.~Schmidt, R.~Widmer, V.~N. Strocov, C.~Felser,
\newblock \emph{Science} \textbf{2020}, \emph{369} 179.

\bibitem{Yao2020}
M.~Yao, K.~Manna, Q.~Yang, A.~Fedorov, V.~Voroshnin, B.~Valentin~Schwarze,
  J.~Hornung, S.~Chattopadhyay, Z.~Sun, S.~N. Guin, J.~Wosnitza, H.~Borrmann,
  C.~Shekhar, N.~Kumar, J.~Fink, Y.~Sun, C.~Felser,
\newblock \emph{Nat. Commun.} \textbf{2020}, \emph{11} 2033.

\bibitem{Sessi2020}
P.~Sessi, F.-R. Fan, F.~K\"{u}ster, K.~Manna, N.~B.~M. Schro\"{o}ter, J.-R. Ji,
  S.~Stolz, J.~A. Krieger, D.~Pei, T.~K. Kim, P.~Dudin, C.~Cacho, R.~Widmer,
  H.~Borrmann, W.~Shi, K.~Chang, Y.~Sun, C.~Felser, S.~S.~P. Parkin,
\newblock \emph{Nat. Commun.} \textbf{2020}, \emph{11} 3507.

\bibitem{Krieger2024}
J.~A. Krieger, S.~Stolz, I.~Robredo, K.~Manna, E.~C. McFarlane, M.~Date, E.~B.
  Guedes, J.~H. Dil, C.~Shekhar, H.~Borrmann, Q.~Yang, M.~Lin, V.~N. Strocov,
  M.~Caputo, B.~Pal, M.~D. Watson, T.~K. Kim, C.~Cacho, F.~Mazzola, J.~Fujii,
  I.~Vobornik, S.~S.~P. Parkin, B.~Bradlyn, C.~Felser, M.~G. Vergniory,
  N.~B.~M. Schr\"{o}ter,
\newblock \emph{Nat. Commun.} \textbf{2024}, \emph{15} 3720.

\bibitem{Zhang2018}
T.~Zhang, Z.~Song, A.~Alexandradinata, H.~Weng, C.~Fang, L.~Lu, Z.~Fang,
\newblock \emph{Phys. Rev. Lett.} \textbf{2018}, \emph{120} 016401.

\bibitem{Miao2018}
H.~Miao, T.~T. Zhang, L.~Wang, D.~Meyers, A.~H. Said, Y.~L. Wang, Y.~G. Shi,
  H.~M. Weng, Z.~Fang, M.~P.~M. Dean,
\newblock \emph{Phys. Rev. Lett.} \textbf{2018}, \emph{121} 035302.

\bibitem{Li2022}
G.~Li, H.~Yang, P.~Jiang, C.~Wang, Q.~Cheng, S.~Tian, G.~Han, C.~Shen, X.~Lin,
  H.~Lei, W.~Ji, Z.~Wang, H.-J. Gao,
\newblock \emph{Nat. Commun.} \textbf{2022}, \emph{13} 2914.

\bibitem{Rao2023}
Z.~Rao, Q.~Hu, S.~Tian, Q.~Qu, C.~Chen, S.~Gao, Z.~Yuan, C.~Tang, W.~Fan,
  J.~Huang, Y.~Huang, L.~Wang, L.~Zhang, F.~Li, K.~Wang, H.~Yang, H.~Weng,
  T.~Qian, J.~Xu, K.~Jiang, H.~Lei, Y.-J. Sun, H.~Ding,
\newblock \emph{Sci. Bull.} \textbf{2023}, \emph{68} 165.

\bibitem{Mardanya2024}
S.~Mardanya, M.~Kargarian, R.~Verma, T.-R. Chang, S.~Chowdhury, H.~Lin,
  A.~Bansil, A.~Agarwal, B.~Singh,
\newblock \emph{Phys. Rev. Mater.} \textbf{2024}, \emph{8} L091801.

\bibitem{Carnicom2018}
E.~M. Carnicom, W.~Xie, T.~Klimczuk, J.~J. Lin, K.~G{\'o}rnicka, Z.~Sobczak,
  N.~P. Ong, R.~J. Cava,
\newblock \emph{Sci. Adv.} \textbf{2018}, \emph{4} eaar7969.

\bibitem{Sun2015}
Z.~X. Sun, M.~Enayat, A.~Maldonado, C.~Lithgow, E.~Yelland, D.~C. Peets,
  A.~Yaresko, A.~P. Schnyder, P.~Wahl,
\newblock \emph{Nat. Commun.} \textbf{2015}, \emph{6} 6633.

\bibitem{yuan2006}
H.~Q. Yuan, D.~F. Agterberg, N.~Hayashi, P.~Badica, D.~Vandervelde, K.~Togano,
  M.~Sigrist, M.~B. Salamon,
\newblock \emph{Phys. Rev. Lett.} \textbf{2006}, \emph{97} 017006.

\bibitem{nishiyama2007}
M.~Nishiyama, Y.~Inada, G.-q. Zheng,
\newblock \emph{Phys. Rev. Lett.} \textbf{2007}, \emph{98} 047002.

\bibitem{Karki2010}
A.~B. Karki, Y.~M. Xiong, I.~Vekhter, D.~Browne, P.~W. Adams, D.~P. Young,
  K.~R. Thomas, J.~Y. Chan, H.~Kim, R.~Prozorov,
\newblock \emph{Phys. Rev. B} \textbf{2010}, \emph{82} 064512.

\bibitem{Amon2018}
A.~Amon, E.~Svanidze, R.~Cardoso-Gil, M.~N. Wilson, H.~Rosner, M.~Bobnar,
  W.~Schnelle, J.~W. Lynn, R.~Gumeniuk, C.~Hennig, G.~M. Luke, H.~Borrmann,
  A.~Leithe-Jasper, Y.~Grin,
\newblock \emph{Phys. Rev. B} \textbf{2018}, \emph{97} 014501.

\bibitem{Khasanov2020b}
R.~Khasanov, R.~Gupta, D.~Das, A.~Amon, A.~Leithe-Jasper, E.~Svanidze,
\newblock \emph{Phys. Rev. Res.} \textbf{2020}, \emph{2} 023142.

\bibitem{Tsvyashchenko2016}
A.~Tsvyashchenko, V.~Sidorov, A.~Petrova, L.~Fomicheva, I.~Zibrov,
  V.~Dmitrienko,
\newblock \emph{J. Alloy. Compd.} \textbf{2016}, \emph{686} 431.

\bibitem{Joshi2015}
B.~Joshi, A.~Thamizhavel, S.~Ramakrishnan,
\newblock \emph{J. Phys.: Conf. Ser.} \textbf{2015}, \emph{592} 012069.

\bibitem{Smidman2017}
M.~Smidman, M.~B. Salamon, H.~Q. Yuan, D.~F. Agterberg,
\newblock \emph{Rep. Prog. Phys.} \textbf{2017}, \emph{80} 036501.

\bibitem{Gao2022}
Z.~S. Gao, X.-J. Gao, W.-Y. He, X.~Y. Xu, T.~K. Ng, K.~T. Law,
\newblock \emph{Quantum Front.} \textbf{2022}, \emph{1} 3.

\bibitem{Lv2019}
B.~Q. Lv, Z.-L. Feng, J.-Z. Zhao, N.~F.~Q. Yuan, A.~Zong, K.~F. Luo, R.~Yu,
  Y.-B. Huang, V.~N. Strocov, A.~Chikina, A.~A. Soluyanov, N.~Gedik, Y.-G. Shi,
  T.~Qian, H.~Ding,
\newblock \emph{Phys. Rev. B} \textbf{2019}, \emph{99} 241104.

\bibitem{Lee2021}
C.~Lee, C.~Yoon, T.~Kim, S.~B. Chung, H.~Min,
\newblock \emph{Phys. Rev. B} \textbf{2021}, \emph{104} L241115.

\bibitem{Kamaeva2022}
L.~V. Kamaeva, M.~V. Magnitskaya, A.~A. Suslov, A.~V. Tsvyashchenko, N.~M.
  Chtchelkatchev,
\newblock \emph{J. Phys.: Condens. Matter} \textbf{2022}, \emph{34} 424001.

\bibitem{Okamoto2020}
Y.~Okamoto, R.~Mizutani, Y.~Yamakawa, H.~Takatsu, H.~Kageyama, K.~Takenaka,
\newblock \emph{JPS Conf. Proc.} \textbf{2020}, \emph{29} 011001.

\bibitem{Mizutani2019}
R.~Mizutani, Y.~Okamoto, H.~Nagaso, Y.~Yamakawa, H.~Takatsu, H.~Kageyama,
  S.~Kittaka, Y.~Kono, T.~Sakakibara, K.~Takenaka,
\newblock \emph{J. Phys. Soc. Jpn.} \textbf{2019}, \emph{88} 093709.

\bibitem{chaikin1995principles}
P.~M. Chaikin, T.~C. Lubensky, T.~A. Witten,
\newblock \emph{Principles of Condensed Matter Physics}, volume~10,
\newblock Cambridge University Press, Cambridge, \textbf{1995}.

\bibitem{dresselhaus2007group}
M.~S. Dresselhaus, G.~Dresselhaus, A.~Jorio,
\newblock \emph{Group Theory: Application to the Physics of Condensed Matter},
\newblock Springer Science \& Business Media, Berlin, \textbf{2007}.

\bibitem{Singh2019b}
D.~Singh, A.~D. Hillier, R.~P. Singh,
\newblock \emph{Phys. Rev. B} \textbf{2019}, \emph{99} 134509.

\bibitem{Chevalier1982}
B.~Chevalier, P.~Lejay, A.~Cole, M.~Vlasse, J.~Etourneau,
\newblock \emph{Solid State Commun.} \textbf{1982}, \emph{41} 801.

\bibitem{Braun1984}
H.~F. Braun,
\newblock \emph{J. Less-Common Met.} \textbf{1984}, \emph{100} 105.

\bibitem{Yaouanc2011}
A.~Yaouanc, P.~D. de~R\'eotier,
\newblock \emph{Muon Spin Rotation, Relaxation, and Resonance: Applications to
  Condensed Matter},
\newblock Oxford University Press, Oxford, \textbf{2011}.

\bibitem{Amato2024}
A.~Amato, E.~Morenzoni,
\newblock \emph{Introduction to Muon Spin Spectroscopy: Applications to Solid
  State and Material Sciences},
\newblock Springer, Cham, \textbf{2024}.

\bibitem{Blundell2021}
S.~J. Blundell, R.~De~Renzi, T.~Lancaster, F.~L. Pratt, editors,
\newblock \emph{Muon {Spectroscopy}: {An} {Introduction}},
\newblock Oxford University Press, Oxford, \textbf{2021}.

\bibitem{Shang2021b}
T.~Shang, T.~Shiroka,
\newblock \emph{Front. Phys.} \textbf{2021}, \emph{9} 270, and references
  therein.

\bibitem{Ghosh2020b}
S.~K. Ghosh, M.~Smidman, T.~Shang, J.~F. Annett, A.~D. Hillier, J.~Quintanilla,
  H.~Yuan,
\newblock \emph{J. Phys.: Condens. Mat.} \textbf{2020}, \emph{33} 033001.

\bibitem{bauer2012}
E.~Bauer, M.~Sigrist, editors,
\newblock \emph{Non-Centrosymmetric Superconductors}, volume 847,
\newblock Springer Verlag, Berlin, \textbf{2012}.

\bibitem{Bauer2004}
E.~Bauer, G.~Hilscher, H.~Michor, C.~Paul, E.~W. Scheidt, A.~Gribanov,
  Y.~Seropegin, H.~No{\"e}l, M.~Sigrist, P.~Rogl,
\newblock \emph{Phys. Rev. Lett.} \textbf{2004}, \emph{92} 027003.

\bibitem{Shang2020a}
T.~Shang, M.~Smidman, A.~Wang, L.-J. Chang, C.~Baines, M.~K. Lee, Z.~Y. Nie,
  G.~M. Pang, W.~Xie, W.~B. Jiang, M.~Shi, M.~Medarde, T.~Shiroka, H.~Q. Yuan,
\newblock \emph{Phys. Rev. Lett.} \textbf{2020}, \emph{124} 207001.

\bibitem{schnyder2015topological}
A.~P. Schnyder, P.~M. Brydon,
\newblock \emph{J. Phys.: Condens. Matter} \textbf{2015}, \emph{27} 243201.

\bibitem{Jorda1982}
J.~Jorda, M.~Ishikawa, J.~Muller,
\newblock \emph{J. Less-Common Met.} \textbf{1982}, \emph{85} 27.

\bibitem{Pereti2023}
C.~Pereti, K.~Bernot, T.~Guizouarn, F.~Laufek, A.~Vymazalov\'{a}, L.~Bindi,
  R.~Sessoli, D.~Fanelli,
\newblock \emph{npj Comput. Mater.} \textbf{2023}, \emph{9} 71.

\bibitem{Hulliger1963}
F.~Hulliger,
\newblock \emph{Nature} \textbf{1963}, \emph{198} 382.

\bibitem{Kudo2018}
K.~Kudo, T.~Takeuchi, H.~Ota, Y.~Saito, S.-y. Ayukawa, K.~Fujimura, M.~Nohara,
\newblock \emph{J. Phys. Soc. Jpn.} \textbf{2018}, \emph{87} 073708.

\bibitem{Sdachi2024}
T.~Adachi, T.~Ogawa, Y.~Komiyama, T.~Sumura, Y.~Saito-Tsuboi, T.~Takeuchi,
  K.~Mano, K.~Manabe, K.~Kawabata, T.~Imazu, A.~Koda, W.~Higemoto, H.~Okabe,
  J.~G. Nakamura, T.~U. Ito, R.~Kadono, C.~Baines, I.~Watanabe, Y.~Imai,
  J.~Goryo, M.~Nohara, K.~Kudo,
\newblock Spontaneous magnetic field and disorder effects in
  {BaPtAs}$_{1-x}${Sb}$_{x}$ with honeycomb network, \textbf{2024},
\newblock ArXiv:2409.05266.

\bibitem{Chen2023}
X.-F. Chen, W.~Luo, T.-F. Fang, Y.~Paltiel, O.~Millo, A.-M. Guo, Q.-F. Sun,
\newblock \emph{Phys. Rev. B} \textbf{2023}, \emph{108} 035401.

\bibitem{Wan2024}
Z.~Wan, G.~Qiu, H.~Ren, Q.~Qian, Y.~Li, D.~Xu, J.~Zhou, J.~Zhou, B.~Zhou,
  L.~Wang, T.-H. Yang, Z.~Sofer, Y.~Huang, K.~L. Wang, X.~Duan,
\newblock \emph{Nature} \textbf{2024}, \emph{632}, 8023 69.

\bibitem{Amato2017}
A.~Amato, H.~Luetkens, K.~Sedlak, A.~Stoykov, R.~Scheuermann, M.~Elender,
  A.~Raselli, D.~Graf,
\newblock \emph{Rev. Sci. Instrum.} \textbf{2017}, \emph{88} 093301.

\bibitem{Kresse1996kl}
G.~Kresse, J.~Furthm\"uller,
\newblock \emph{Phys. Rev. B} \textbf{1996}, \emph{54} 11169.

\bibitem{Kresse1996vk}
G.~Kresse, J.~Furthm{\"u}ller,
\newblock \emph{Comput. Mater. Sci.} \textbf{1996}, \emph{6} 15.

\bibitem{Kresse1999wc}
G.~Kresse, D.~Joubert,
\newblock \emph{Phys. Rev. B} \textbf{1999}, \emph{59} 1758.

\bibitem{Blochl1994zz}
P.~E. Bl\"ochl,
\newblock \emph{Phys. Rev. B} \textbf{1994}, \emph{50} 17953.

\bibitem{Perdew1996iq}
J.~P. Perdew, K.~Burke, M.~Ernzerhof,
\newblock \emph{Phys. Rev. Lett.} \textbf{1996}, \emph{77} 3865.

\bibitem{pizzi_wannier90_2020}
G.~Pizzi, V.~Vitale, R.~Arita, S.~Bl\"{u}gel, F.~Freimuth, G.~G\'{e}ranton,
  M.~Gibertini, D.~Gresch, C.~Johnson, T.~Koretsune, J.~Iba\~{n}ez Azpiroz,
  H.~Lee, J.-M. Lihm, D.~Marchand, A.~Marrazzo, Y.~Mokrousov, J.~I. Mustafa,
  Y.~Nohara, Y.~Nomura, L.~Paulatto, S.~Ponc\'{e}, T.~Ponweiser, J.~Qiao,
  F.~Th\"{o}le, S.~S. Tsirkin, M.~Wierzbowska, N.~Marzari, D.~Vanderbilt,
  I.~Souza, A.~A. Mostofi, J.~R. Yates,
\newblock \emph{J. Phys.: Condens. Matter} \textbf{2020}, \emph{32} 165902.

\bibitem{wu_wanniertools_2018}
Q.~Wu, S.~Zhang, H.-F. Song, M.~Troyer, A.~A. Soluyanov,
\newblock \emph{Comput. Phys. Commun.} \textbf{2018}, \emph{224} 405.

\bibitem{Zhu2008}
X.~Zhu, H.~Yang, L.~Fang, G.~Mu, H.-H. Wen,
\newblock \emph{Supercond. Sci. Technol.} \textbf{2008}, \emph{21} 105001.

\bibitem{Suter2012}
A.~Suter, B.~M. Wojek,
\newblock \emph{Phys. Procedia} \textbf{2012}, \emph{30} 69.

\end{thebibliography}
\clearpage

%==== figure =============================%
%=== end figure ==========================%
\clearpage

%==== figure =============================%
%=== end figure ==========================%
\clearpage

%=== begin figure ==========================%
%=== end figure ==========================% 
\clearpage

%=== begin figure ==========================%
%=== end figure ==========================% 
\clearpage

%=== begin figure ==========================%
%=== end figure ==========================%  
\clearpage

% Essays at most 5000 words long from https://cn.leifu.whu.edu.cn/info/1043/1498.htm

% Please provide Biographies and photos for Essays, Feature Articles, Progress Reports, Reviews, and Perspectives for those authors who should be highlighted  
% These should be at most 100 words long
% For other article types this section can be removed
% Photographs should be 40mm broad and 50 mm high

%\begin{figure}
  %\includegraphics{bio-placeholder.jpg}
  %\caption*{Biography}
%\end{figure}

% Table of contents entry should be 50 - 60 words long
% Image should be 55 mm broad and 50 mm high or 110 mm broad and 20 mm high

%\begin{figure}
%\textbf{Table of Contents}\\
%\medskip
  %\includegraphics{toc-image.png}
  %\medskip
  %\caption*{ToC Entry}
%\end{figure}

\end{document}